%% file: main.tex
\documentclass[
    prd, aps, twocolumn, nofootinbib, showpacs, superscriptaddress
]{revtex4-2}
\usepackage[colorlinks=true,citecolor=green,urlcolor=blue,linkcolor=blue]{hyperref}
\usepackage[utf8]{inputenc}
\usepackage{graphicx,mathtools,xcolor,latexsym,rotating,soul}
\usepackage[T1]{fontenc}
\usepackage{aas_macros,natbib,tikz}
\usepackage{amsmath, amssymb, amsthm, amsfonts,breqn}
\usetikzlibrary{decorations.pathmorphing}

\newtheorem{theorem}{Theorem}

\newtheorem{lemma}[theorem]{Lemma}
\newtheoremstyle{named}{}{}{\itshape}{}{\bfseries}{.}{.5em}{\thmnote{#3 }#1}
\theoremstyle{named}
\newtheorem*{namedtheorem}{}
\newcommand{\F}{\mathcal{F}}

\newcommand{\C}{\mathcal{C}}
\newcommand{\Oo}{\mathcal{O}}

\newcommand{\omegaH}{\Omega_{\mathcal{H}}}

\definecolor{orange}{rgb}{1.0, 0.5, 0.0}

\makeatletter
\let\cat@comma@active\@empty
\makeatother
\begin{document}
\title{How Do Axisymmetric Black Holes Grow Monopole and Dipole Hair?}
\author{Abhishek Hegade K R}
\affiliation{Illinois Center for Advanced Studies of the Universe, Department of Physics, University of Illinois at Urbana-Champaign, Urbana, IL 61801, USA}

\author{Elias R. Most}
\affiliation{Princeton Center for Theoretical Science, Princeton University, Princeton, NJ 08544, USA}
\affiliation{Princeton Gravity Initiative, Princeton University, Princeton, NJ 08544, USA}
\affiliation{School of Natural Sciences, Institute for Advanced Study, Princeton, NJ 08540, USA}

\author{Jorge Noronha}
\affiliation{Illinois Center for Advanced Studies of the Universe, Department of Physics,
University of Illinois at Urbana-Champaign, Urbana, IL 61801, USA}

\author{Helvi Witek}
\affiliation{Illinois Center for Advanced Studies of the Universe, Department of Physics, University of Illinois at Urbana-Champaign, Urbana, IL 61801, USA}

\author{Nicol\'as Yunes}
\affiliation{Illinois Center for Advanced Studies of the Universe, Department of Physics, University of Illinois at Urbana-Champaign, Urbana, IL 61801, USA}

\begin{abstract}
We study the dynamical formation of scalar monopole and dipole hair in scalar Gauss-Bonnet theory and dynamical Chern-Simons theory.
We prove that the spherically-symmetric mode of the dipole hair is completely determined by the product of the mass of the spacetime and the value of the monopole hair.
We then show that the dynamics of the $\ell=1$ mode of the dipole hair is intimately tied to the appearance of the event horizon during axisymmetric collapse, which results in the radiation of certain modes that could have been divergent in the future of the collapse.
We confirm these analytical predictions by simulating the gravitational collapse of a rapidly rotating
neutron star
in the decoupling limit, both in scalar Gauss-Bonnet and dynamical Chern-Simons theory.
Our results, combined with those of Ref.~\cite{R:2022cwe}, provide a clear physical picture of the dynamics of scalar monopole and dipole radiation in axisymmetric and spherical gravitational collapse in these theories.
\end{abstract}
\maketitle
\section{Introduction}
General relativity (GR) predicts that black holes (BHs) represent one of the simplest macroscopic objects in nature, characterized solely by their mass, spin and charge~\cite{1967-Israel,1968-Israel,1971-Carter,1972-Hawking-No-Hair}. The detection of gravitational waves by the advanced Laser Interferometer Gravitational-Wave Observatory (LIGO) and Virgo~\cite{LVK-detection,2019-GWTC-1,2021-GWTC-2,LIGOScientific:2021djp} 
now allows for tests of the hypothesis that the astrophysical BHs of Nature are truly as GR predicts~\cite{TheLIGOScientific:2016src,Yunes_2016,Abbott:2018lct,Nair_2019,Perkins_2021,LIGOScientific:2021sio,Lyu_2022}. 
Motivated by these tests, several non-GR modified gravity theories have been proposed in the literature~\cite{Yunes_2013,Berti:2015itd}. 
Some of these theories allow for qualitatively different BHs, endowed with additional ``hair'' that is needed to fully characterize them~(see Ref.~\cite{Yunes_2013,Berti:2015itd,herdeiro2018asymptotically} for a review). 
One class of theories that admits such hairy solutions are models in which a scalar or pseudo-scalar field couples non-minimally  to a curvature invariant.
Modified theories that fall in this class include scalar Gauss-Bonnet (SGB) theory~\cite{1996-Kanti,Mignemi:1992nt}, dynamical Chern-Simons (DCS) gravity~\cite{2003-Jackiw-Pi,2009-Alexander-Yunes}, and modified quadratic theories of gravity in general~\cite{Kanti:1995cp,Yunes_2013,Cano:2021rey}. 
The non-minimal coupling to the Gauss-Bonnet invariant in SGB theory and to the Pontryagin density in DCS gravity appear naturally in the low-energy limits
of heterotic string theory~\cite{Zwiebach:1985uq,Gross:1986mw,Kanti:1995cp,Taveras:2008yf,2009-Alexander-Yunes,Cano:2021rey}
and, more generally,  
in effective field theories that include a real scalar field~\cite{Weinberg_2008}. 
We here focus on theories that smoothly reduce to GR in the small coupling limit. For non-perturbative, non-minimal coupling functions between the scalar field and curvature invariants we refer the interested reader to the recent review in Ref.~\cite{Doneva:2022ewd}.

The presence of a scalar or pseudo-scalar field leads to scalar radiation
in BH and neutron star (NS) binary systems which can have observable effects, such as dephasing of gravitational waves~\cite{2012-Yagi-Stein-Yunes-Tanaka,Okounkova:2017yby,Witek:2018dmd,Okounkova:2019zep,Ripley-East-2021,Shiralilou:2020gah,Shiralilou:2021mfl,CCZ4-2022}. 
The presence of this observable effect and the possibility of constraining these theories have led to an enormous amount of work concerning the properties of these theories.
Several studies have explored the space of BH solutions in static and slowly rotating BH solutions in the small coupling approximation~\cite{1996-Kanti,2011-Stein-Yunes,Sotiriou-Zhou-2014,Sotiriou:2014PRD,Ayzenberg:2014aka,2011-Pani-Macedo-Cardoso,2015-Maselli} for SGB theory and in~\cite{Yunes:2009hc,Yagi:2012ya,2011-Pani-Macedo-Cardoso,2011-Stein-Yunes,2011-Pani-Macedo-Cardoso} for DCS gravity, while numerical solutions for static and stationary BHs were calculated in~\cite{Sotiriou-Zhou-2014,Sotiriou:2014PRD,2021-Sullivan} for SGB theory and in~\cite{2018-Delsate} for DCS gravity.
Much effort has also been devoted towards understanding the dynamics of BH binary inspirals and BH-NS binary systems using analytical techniques, such as
post-Newtonian theory~\cite{2012-Yagi-Stein-Yunes-Tanaka,Julie:2019sab,Shiralilou:2020gah,Shiralilou:2021mfl}, using the tools of numerical relativity~\cite{Ripley:2022cdh,Benkel-2016,Benkel_2017,Witek:2018dmd,Okounkova:2017yby,Okounkova:2019zep,Okounkova:2020rqw,Ripley-East-2021,Ripley-Pretorius-2019,Ripley-Pretorius-2020(1),Ripley-Pretorius-2020,East-2022,CCZ4-2022,Silva:2020omi,Corman:2022xqg}  
and using the tools of 
black hole perturbation theory~\cite{Molina:2010fb,Blazquez-Salcedo:2016enn,Bryant:2021xdh,Wagle:2021tam,Pierini:2021jxd,Hui:2021cpm,Cano:2021myl,Li:2022pcy}. 
More recently, well-posedness~\cite{Kovacs_2020_PRL,Reall_2021} and the loss of predictivity have also been explored in SGB theory~\cite{Ripley-Pretorius-2019,Ripley-Pretorius-2020(1),Ripley-Pretorius-2020,Ripley-East-2021,ahkr-ripley-yunes-2022}.

Although impressive, these studies were geared towards providing a phenomenological understanding of
scalar and gravitational radiation and in exploring the breakdown of predictivity in these theories.
In Ref.~\cite{R:2022cwe} (Paper-I, henceforth) we took the first steps towards providing a
theoretical understanding of
the
scalar field dynamics during 
gravitational collapse.
While Paper-I focused on spherical symmetry, in the present paper we describe the physical mechanism 
behind the emission of scalar radiation during gravitational collapse
in axi-symmetry.

The long range dynamics of the scalar field are quantified by studying its far field behaviour.
In the exterior spacetime, far away from a compact object, the scalar field can be expanded in powers of $1/r$, where $r$ is a suitable distance measure from the compact object. 
The coefficient of the leading $1/r$ term in the far-field expansion of the scalar field is called the ``monopole hair'' and the sub-leading $1/r^{2}$ term is called the ``dipole hair.''
The monopole hair and the dipole hair can be further classified by their angular modes. 
The monopole hair contains only an $\ell=0$ mode, while the dipole hair contains both an $\ell=0$ and an $\ell=1$ mode. 

The monopole and dipole hair in modified theories, such as SGB gravity and DCS gravity, display  interesting phenomena during dynamical gravitational interactions, such as during the collapse of a NS into a BH.
In SGB gravity, the monopole hair of a NS spacetime is zero~\cite{2012-Yagi-Stein-Yunes-Tanaka,2016-Yagi-Stein-Yunes}, while the monopole hair for a BH spacetime is non-zero
and is related to the surface gravity and the topology of the bifurcation 2-sphere~\cite{Prabhu-Stein-2018}. 
Therefore, the monopole hair in SGB theory must grow during gravitational collapse from a NS spacetime to a BH spacetime.
On the other hand, in DCS theory, the monopole hair vanishes for NS and BH spacetimes~\cite{Wagle:2018tyk,2016-Yagi-Stein-Yunes,2012-Yagi-Stein-Yunes-Tanaka}.
Therefore, the dynamics of scalar radiation in DCS theory is controlled by dipole radiation.
In Paper-I, we 
showed that the growth of hair during spherically-symmetric gravitational collapse is related to the appearance of the EH, which results in the radiation of certain homogeneous modes that can be present in a NS spacetime but cannot be present in a BH spacetime.
In this work, we show that this analysis also carries over to axisymmetric gravitational collapse and to the dynamics of scalar dipole radiation for both SGB theory and DCS theory.

First, we show that there is a remarkably simple relationship between monopole hair and dipole hair.
In particular, we show that the $\ell=0$ mode of the dipole hair 
is equal to the product of the mass of spacetime and the monopole hair.
Moreover, in a spherically-symmetric spacetime, the dipole hair contains only the $\ell=0$ mode and we can readily extend
our results in Paper-I from monopole to dipole hair.
In fact, our results predict that the dipole hair must also grow
during
gravitational collapse in SGB theory,
and that the growth of dipole hair is related to the radiation of certain divergent homogeneous modes, just as in the monopole case.
With this observation, we are now able to  provide a clear description of the formation
of monopole and dipole hair in spherically-symmetric gravitational collapse in SGB theory.

Next, we explore the dynamics of monopole and dipole scalar radiation in axisymmetric spacetimes in SGB theory and DCS theory.
We first provide a general Green's formula to calculate the $\ell=1$ mode of the dipole hair.
Using this formula, we show that $\ell=1$ mode of the dipole hair vanishes in SGB theory due to parity.
This means that understanding the dynamics of monopole hair in SGB theory allows one to fully understand the dynamics of dipole hair in axisymmetric gravitational collapse at early and late times.
We then explore the dynamics of DCS theory in axisymmetric gravitational collapse.
Our analytical formula shows that the vanishing of monopole hair in DCS theory implies that the $\ell=0$ mode of the dipole hair also vanishes.
This means that scalar radiation in DCS theory must be driven by the $\ell=1$ mode of the dipole hair.
Using analytical calculations in the
slow-rotation
approximation, we show that the $\ell=1$ mode of dipole hair in a NS spacetime in DCS theory contains homogeneous modes that can be divergent if present in a BH spacetime.
Therefore, these modes must be radiated during dynamical gravitational collapse, using the same physical mechanism we described in Paper-I.
In particular, the appearance of the event horizon (EH) leads to the radiation of these modes.

Finally, we confirm all our analytical predictions by simulating the axisymmetric gravitational collapse of a rapidly-rotating NS in the decoupling limit in both SGB theory and DCS theory.
We find good agreement with our analytical predictions at early and late times. We also show that the appearance of the EH leads to strong scalar radiation that can be correlated with the dynamics of scalar monopole and dipole radiation.

The remainder of the paper is organized as follows. 
In Sec.~\ref{sec:Monopole-Dipole} we present the field equations and classify the monopole and dipole hair in spherically-symmetric and axi-symmetric spacetimes.
The main analytical results are presented in this section.
In Sec.~\ref{sec:dynamics} we study the dynamics of the scalar field in SGB gravity and DCS gravity in the decoupling limit. We first present analytical solutions in the slow-rotation approximation and then present numerical results for the collapse of a rapidly rotating NS to a BH spacetime.
Our conclusions and directions for future work are presented in Sec.~\ref{sec:Conclusions}.
Our metric signature is $(-,+,+,+)$ and we set $G=1=c$ throughout the paper.

\section{Monopole and Dipole Hair in Stationary Spacetimes}\label{sec:Monopole-Dipole}
In this section, we describe our analytical results in detail.
First, we introduce the field equations in Sec.~\ref{sec:Field-eqs}.
We then provide the formula for the scalar dipole hair in terms of the monopole hair in Sec.~\ref{sec:Monopole-Dipole-Spherically-Symmetric} for spherically-symmetric spacetimes and in Sec.~\ref{sec:dipole-axisymmetric} for axi-symmetric spacetimes. 
Finally, we discuss how these results can be applied to specific theories of gravity, such as SGB theory and DCS theory in Sec.~\ref{sec:App-SGB-DCS}.
\subsection{Action and Field Equations}\label{sec:Field-eqs}
We study a general class of theories with a scalar field $\Phi$ coupled non-minimally to gravity through a curvature scalar $\F$. 
The strength of this coupling is quantified by the coupling constant $\epsilon$. The action for this class of theories is given by
\begin{align}\label{eq:Action}
     S &= \int d^4x \sqrt{-g} \left( \frac{1}{16\pi} R + \epsilon\, \Phi \,\mathcal{F}(g,\partial g,\partial^2 g,\ldots) \right.\nonumber \\
     &- \left. \frac{1}{2} \left( \nabla_{\mu}\Phi \nabla^{\mu}\Phi  \right)\right) + S_{\text{matter}}\,,
\end{align}
where $g$ denotes the determinant of the spacetime metric, $R$ is the Ricci scalar, $\nabla$ is the covariant derivative, and $S_{\text{matter}}$ is the action for the matter fields, which is assumed to be minimally coupled to gravity and independent of the scalar field $\Phi$.

Let us map the action above into the specific modified theories of gravity considered in this work. For example, we can recover the action for
shift-symmetric
SGB theory by replacing
\begin{align}
    \epsilon &\to \alpha_{\text{SGB}} \, \nonumber,\\
    \F &\to \mathcal{R}_{\text{GB}} =R^2 - 4R_{\mu\nu}R^{\mu\nu} + R_{\alpha\beta\gamma\delta}R^{\alpha\beta\gamma\delta}\nonumber,
\end{align}
in Eq.~\eqref{eq:Action}.
The scalar field $\Phi = \phi_{\text{SGB}}$ now represents the dilaton field.
Similarly, we can recover DCS gravity by replacing
\begin{align}
    \epsilon &\to \alpha_{\text{DCS}}/4 \, \nonumber,\\
    \F &\to R_{\beta\alpha\gamma\delta}{}^{*}R^{\alpha\beta\gamma\delta}\nonumber,
\end{align}
in the action.
The scalar field $\Phi = \theta_{\text{DCS}}$ now represents the axion pseudoscalar.

The equations of motion for the action in Eq.~(\ref{eq:Action}) are given by
\begin{align}\label{eq:EOM}
    G_{\mu\nu} + 16\pi \,\epsilon\, \mathcal{C}_{{\mu\nu}} &= 8\pi \left(T_{\mu\nu}^{\Phi} +  T_{\mu\nu}^{\text{matter}} \right) \nonumber, \\
    \Box{\Phi} + \epsilon \,\mathcal{F} = 0. 
\end{align}
The stress energy tensor for $\Phi$ is 
\begin{equation}\label{eq:T-phi}
    T_{\mu\nu}^{\Phi} =  \nabla_{\mu}\Phi\nabla_{\nu}\Phi - \frac{g_{\mu\nu}}{2}\left( \nabla^{\delta}\Phi\nabla_{\delta}\Phi\right),
\end{equation}
while the tensor $\mathcal{C}_{\mu\nu}$ is given by
\begin{equation}\label{eq:C-Tensor}
    \mathcal{C}_{\mu\nu} := \frac{1}{\sqrt{-g}}\frac{\delta}{\delta g^{\mu\nu}} \int d^4x \sqrt{-g}\, \Phi \mathcal{F}(g,\partial g,\partial^2 g,\ldots)  .
\end{equation}
The explicit expressions for
the $\C_{\mu\nu}$
tensor for modified quadratic gravity theories, such as SGB theory and DCS theory, can be found in Ref.~\cite{Yunes_2013}.
Finally, the matter stress energy tensor is defined by
\begin{equation}
    T_{\mu\nu}^{\text{matter}} = \frac{-2}{\sqrt{-g}}\frac{\delta}{\delta g^{\mu\nu}} S_{\text{matter}}\, .
\end{equation}

\subsection{Monopole and Dipole Hair in spherically-symmetric Spacetimes}\label{sec:Monopole-Dipole-Spherically-Symmetric}
In this subsection, we analyze the behaviour of the scalar field near spatial infinity in static, spherically-symmetric, and asymptotically flat spacetimes. 
We begin by introducing the ingoing null coordinate system $x^{\mu} = (v,r,\theta,\phi)$ on a spherically-symmetric spacetime. 
The line element in this coordinate system is given by
\begin{equation}\label{eq:G-Null-MC}
    ds^2 = -D(r)dv^2 + 2dvdr + K(r) d\Omega^2\,,
\end{equation}
where $K(r)^{-1}$ denotes the Gaussian curvature of the 2-sphere parameterised by $(\theta,\phi)$, while $d\Omega^2$ denotes the line element on the 2-sphere.
The metric introduced in Eq.~\eqref{eq:G-Null-MC} is valid for both NS and BH spacetimes. For BH spacetimes, the location of the EH, 
$r_{H}$, is defined by the condition
\begin{equation}
    D(r_H) = 0\,.
\end{equation}
The analysis we present below will apply to both BH and NS spacetimes, since we will be analyzing the scalar field equation, Eq.~\eqref{eq:EOM}. However, to simplify the presentation of our results, we will assume that we are in a BH spacetime and that the EH is located at $r=r_H$. One can easily transform the results given below to NS spacetimes by replacing $r_H\to 0$.

We are interested in understanding the asymptotic properties of the scalar field $\Phi$. Near spatial infinity, we assume that the scalar field and the metric variables admit a smooth expansion in powers of $r^{-1}$. Expanding the scalar field in powers of $r^{-1}$ gives
\begin{equation}\label{eq:Asymptotic-Expansion-scalar}
    \Phi(r) = \frac{\mu_1}{r} + \frac{\mu_2}{r^2} + \Oo(r^{-3})\,,
\end{equation}
where $\mu_1$ is  the monopole scalar hair, $\mu_2$ is  the dipole scalar hair, and the ${\cal{O}}$ symbol stands for uncontrolled remainders hereafter. The properties of monopole hair have been investigated for specific theories, such as SGB theory and DCS theory, in Refs.~\cite{2013-Yagi-Stein-Yunes-Tanaka,2016-Yagi-Stein-Yunes,Prabhu-Stein-2018}. In Paper-I we generalised these results and provided a formula for the monopole hair by solving the scalar field equation~\eqref{eq:EOM}
\begin{equation}
    \mu_1 = \epsilon\int_{r_H}^{\infty}\mathcal{F}(x)K(x)\, dx\,.
\end{equation}
For the sake of completeness, we shall re-derive this result below. In the present situation, we are interested in understanding the behaviour of the dipole hair $\mu_2$ in theories where the curvature scalar $\mathcal{F}$ has the following asymptotic behavior near spatial infinity
\begin{equation}\label{eq:Fall-off-curvature-spherical}
    \F \sim  r^{-5}\,.
\end{equation}
This asymptotic expansion is valid for curvature scalars, such as the Gauss-Bonnet invariant, the Kretschmann scalar, and Pontryagin density. 
In fact, since these curvature scalars scale as curvature squared, they have an even stronger asymptotic fall off, decaying as $r^{-6}$.

To study the behaviour of the dipole hair $\mu_2$, we introduce an asymptotically mass centered (AMC) coordinate system. The latter is defined as the coordinate system in which the metric functions $D(r)$ and $K(r)$ have the following asymptotic expansions
\begin{align}
\label{eq:Asymptotic-Expansion-K-D-1}
    D(r) &= 1 -\frac{2M}{r} + \Oo(r^{-2})\,,\\
    \label{eq:Asymptotic-Expansion-K-D-2}
    K(r) &= r^2\left[1+ \Oo(r^{-2})\right]\,.
\end{align}
The quantity $M$ in the equation above denotes the Komar mass of the BH spacetime~\cite{Wald-book}. We note that one can always introduce a coordinate system in which Eqs.~\eqref{eq:Asymptotic-Expansion-K-D-1} and \eqref{eq:Asymptotic-Expansion-K-D-2} are valid by performing suitable translations. We provide a proof of this statement in Appendix~\ref{appendix:AMC-spherical}.
We also note that our notion of an AMC coordinate system is not as strong as the asymptotically Cartesian and mass centered coordinate system (ACMC) introduced by Thorne in Ref.~\cite{Thorne-1980}. ACMC coordinates require that the $\mathcal{O}(r^{-2})$ coefficient of the $g_{vv}$ component of the metric be zero. The AMC coordinate system we introduce does not require this condition.

We now look at the scalar field equations. Using Eq.~\eqref{eq:G-Null-MC} the scalar field equation can be written as
\begin{equation}
    \frac{1}{K(r)} \partial_r \left[K(r) D(r) \partial_r \Phi(r) \right] +\epsilon \; \mathcal{F}=0\,.
\end{equation}
This equation can be integrated as described in Sec. II B of Paper-I,
\begin{align}\label{eq:Spherical-Integrated}
     \partial_r \Phi  &=   -\frac{\epsilon} {D(r)K(r)} \int_{r_H}^{r}\mathcal{F}(x)K(x)\, dx\,, \nonumber\\
     &= -\frac{\epsilon} {D(r)K(r)}\int_{r_H}^{\infty}\mathcal{F}(x)K(x)\, dx \nonumber\\
     &+ \frac{\epsilon}{D(r)K(r)}\int_{r}^{\infty}\mathcal{F}(x)K(x)\, dx\, .
\end{align}
Using Eq.~\eqref{eq:Asymptotic-Expansion-scalar}, we see that the derivative of the scalar field has the following asymptotic expansion
\begin{equation}\label{eq:Asymptotic-Expansion-Der-Scalar}
    \partial_r \Phi = -\frac{\mu_1}{r^2} - \frac{2\mu_2}{r^3} + \Oo(r^{-4})\,.
\end{equation}

Let us now discuss the asymptotic properties of the scalar field by using Eq.~\eqref{eq:Spherical-Integrated}.
Equations~\eqref{eq:Fall-off-curvature-spherical},~\eqref{eq:Asymptotic-Expansion-K-D-1}, ~and~\eqref{eq:Asymptotic-Expansion-K-D-2} tell us that
the second term on the right-hand side of Eq.~\eqref{eq:Spherical-Integrated} has the following asymptotic behavior 
\begin{equation}
    \frac{\epsilon}{D(r)K(r)}\int_{r}^{\infty}\mathcal{F}(x)K(x)\, dx \sim r^{-4}\,.
\end{equation}
With this observation, we see that to determine the value of the monopole and dipole hair of the scalar field we can just look at the asymptotic properties of the first term on the right-hand side of Eq.~\eqref{eq:Spherical-Integrated}.
Using Eqs.~\eqref{eq:Asymptotic-Expansion-K-D-1} and \eqref{eq:Asymptotic-Expansion-K-D-2}, Eq.~\eqref{eq:Spherical-Integrated} can be written as
\begin{align}
    \partial_r \Phi &= 
    -\frac{\epsilon} {D(r)K(r)}\int_{r_H}^{\infty}\mathcal{F}(x)K(x)\, dx + \Oo(r^{-4})\,, \nonumber\\
    &= -\frac{\epsilon\int_{r_H}^{\infty}\mathcal{F}(x)K(x)\, dx}{\left[1- 2Mr^{-1} + \Oo(r^{-2})\right]r^2\left[1 + \Oo(r^{-2})\right]} + \Oo(r^{-4})\,,\nonumber\\
    &= -\left(\!\epsilon\!\int_{r_H}^{\infty}\mathcal{F}(x)K(x)\, dx\right)\left(\frac{1}{r^2} + \frac{2M}{r^3} \right)\! + \! \Oo(r^{-4}) \,.
\end{align}
Comparing this with Eq.~\eqref{eq:Asymptotic-Expansion-Der-Scalar}, we see that the value of the monopole hair and the dipole hair are given by
\begin{align}\label{eq:Monopole-hair-spherical-rederived}
    \mu_1 &= \epsilon\int_{r_H}^{\infty}\mathcal{F}(x)K(x)\, dx\, , \\
    \mu_2 &= M\mu_1 \label{eq:Dipole-hair-spherical}\,.
\end{align}

Equation~\eqref{eq:Monopole-hair-spherical-rederived} gives us a formula for the monopole hair $\mu_1$ in terms of the integral of the curvature invariant $\mathcal{F}$. 
This formula for the monopole hair was also given in Corollary 1.2 of Paper-I.
From  Eq.~\eqref{eq:Dipole-hair-spherical} we see that in AMC coordinates the value of the dipole hair is completely determined by the value of the monopole hair and the mass of the compact object.
Therefore, studying the behaviour of the monopole hair during spherically-symmetric collapse provides us with all the required information to understand the behaviour of the dipole hair of the scalar field.
For example, in Paper-I we studied the gravitational collapse and growth of monopole hair in SGB gravity. With the analysis presented above, we see that the dipole hair must also grow during gravitational collapse and its growth must be correlated with the complete disappearance of the surface of the star inside the EH and the release of scalar radiation~\cite{R:2022cwe}.
\subsection{Monopole and Dipole Hair in Axisymmetric and Circular Spacetimes}\label{sec:dipole-axisymmetric}
We now extend the result of the previous section from spherically-symmetric spacetimes to axisymmetric, circular, and asymptotically flat spacetimes. These are spacetimes in which
\begin{enumerate}
    \item The vector fields, $\partial_t$ and $\partial_\phi$, are killing vectors of the spacetime.
    \item The spacetime is circular, i.e., the 2-surfaces orthogonal to $\partial_t$ and $\partial_\phi$ are integrable.
    \item In addition to the assumptions above, we also assume that the curvature scalar $\mathcal{F}$ falls off asymptotically at least as $r^{-5}$. 
\end{enumerate}
Using the assumptions in (1) and (2), we can introduce Hartle-Thorne type coordinates, $x^{\mu} = (t,r,\theta,\phi)$, on our spacetime~\cite{Hartle-1967}. In these coordinates, the line element takes the following form
\begin{align}\label{eq:Hartle-Thorne-coordinates}
     ds^2 &= -N^2(r,\theta)\, dt^2 + A^2(r,\theta)\, dr^2 \, \nonumber\\
     &+ r^2B^2(r,\theta) \left\{ d\theta^2 + \sin(\theta)^2 \left[d\phi - \omega(r,\theta) dt\right]^2\right\}\,.
\end{align}
We also impose one further assumption, 
\begin{enumerate}
  \setcounter{enumi}{3}
  \item The metric is reflection symmetric, $ g_{\mu\nu}(r,\theta) = g_{\mu\nu}(r,\pi-\theta)$.
\end{enumerate}

We now comment on the motivation behind our assumptions (2), (3), and (4). 
The assumption of circularity is independent from the assumption of stationarity and axisymmetry. From a physical point of view, circular spacetimes are spacetimes for which there is no ``meridional'' motion or currents~\cite{gourgoulhon2011introduction}. 
Therefore, the assumption of circularity is justified from a physical point of view for equilibrium configurations.
For vacuum GR the assumption of circularity follows from the assumption of stationarity and axisymmetry as shown by Papapetrou~\cite{Papapetrou:1966} and Carter~\cite{Carter-1969}. We refer the reader to Chapter 2 of Ref.~\cite{gourgoulhon2011introduction} for a more detailed discussion of circular spacetimes in non-vacuum GR.  Furthermore, Xie et al.~\cite{Xie:2021} showed that if the GR solution is circular then solutions to Eq.~\eqref{eq:EOM} that admit a smooth perturbative expansion in the coupling constant $\epsilon$ are also circular to all orders in perturbation theory\footnote{See Sec III of Ref.~\cite{R:2022cwe} for a detailed discussion of perturbation
theory.}. Assumption (3) does not restrict the class of theories we wish to study since the Gauss-Bonnet, Kretschmann, and Pontryagin curvature scalars all fall off faster than $r^{-5}$ 
near spatial infinity. 
The assumption of reflection symmetry is also physically motivated since we are studying equilibrium configurations. The assumption of reflection symmetry also helps us when setting up AMC coordinates (see Appendix~\ref{appendix:AMC-circular}).

Although the arguments we present below do not depend on whether the spacetime is a BH or a NS spacetime, as in the previous section, we will here assume the spacetime is a BH one to simplify our presentation.
Generalising our definition of AMC coordinates, we say that our coordinates are AMC if the metric function $B(r,\theta)$ has the following asymptotic expansion
\begin{equation}\label{eq:asym-circular-1}
    B = 1 + \Oo(r^{-2})\,.
\end{equation}
In AMC coordinates, the functions $N(r,\theta)$, $A(r,\theta)$, and $\omega(r,\theta)$ have the following asymptotic expansions
\begin{align}
    N &= 1 -\frac{M}{r} + \Oo(r^{-2}) \label{eq:asym-circular-2} \, ,\\
    A &= 1+ \frac{M}{r} + \Oo(r^{-2}) \label{eq:asym-circular-3}\, , \\
    \omega &= \frac{2J}{r^3} + \Oo(r^{-4}) \label{eq:asym-circular-4}\, ,
\end{align}
where $J$ denotes the angular momentum.
Assumptions (1)-(4) imply the existence of AMC coordinates, as we show in Appendix~\ref{appendix:AMC-circular}. 
The proof exploits the assumption that the curvature scalar $\mathcal{F}$ scales as $r^{-5}$ 
near spatial infinity, which implies that Eq.~\eqref{eq:EOM} reduces to the Einstein massless scalar field equation to $\Oo(r^{-4})$.
Solving Eq.~\eqref{eq:EOM} asymptotically to $\Oo(r^{-4})$ shows that one can always employ AMC coordinates.

We now study the behaviour of the scalar monopole and dipole hair.
The scalar field has the following asymptotic expansion
\begin{equation}\label{eq:Asymptotic-Axi-Scalar}
    \Phi = \frac{\mu_1(\theta)}{r} + \frac{\mu_2(\theta)}{r^2} + \Oo(r^{-3})\,,
\end{equation}
where $\mu_1(\theta)$ and $\mu_2(\theta)$ are the monopole and dipole hair, respectively. 
The equation of motion for the scalar field is given by
\begin{equation}\label{eq:E-Phi}
    E_{\Phi} := \Box{\Phi(r,\theta)}+ \epsilon \,\mathcal{F}(r,\theta) = 0\,.
\end{equation}
\subsubsection{Relation between \texorpdfstring{$\mu_1(\theta)$}{Lg} and \texorpdfstring{$\mu_2(\theta)$}{Lg}}
We now establish a relation between the dipole hair and the monopole hair analogous to the one given in Eq.~\eqref{eq:Dipole-hair-spherical}. 
To do this, we plug in the asymptotic expansion of the metric variables and the scalar field in Eqs.~\eqref{eq:asym-circular-1}-\eqref{eq:Asymptotic-Axi-Scalar} into $E_{\Phi}$ and expand asymptotically to obtain,
\begin{equation}\label{eq:asymptotic-eqscalar}
    E_{\Phi} = \frac{f_3(\theta)}{r^3} + \frac{f_4(\theta)}{r^4} + \Oo(r^{-5}) =0\,,
\end{equation}
where
\begin{align}\label{eq:f_3}
    f_3(\theta) &= \cot (\theta ) \frac{d\, \mu _1(\theta )}{d\theta}
    +
    \frac{d^2 \mu _1(\theta )}{d\theta^2}\, ,\\
    \label{eq:f_4}
    f_4(\theta) &= \cot (\theta ) \frac{d\, \mu _2(\theta )}{d\theta}
    +
    \frac{d^2 \mu _2(\theta )}{d\theta^2}+2 \,\mu _2(\theta )-2 M \mu _1(\theta ) \, ,
\end{align}
Since we are at spatial infinity, $f_3(\theta)$ and $f_4(\theta)$ must be equal to zero. This gives us two differential equations for the variables $\mu_1(\theta)$ and $\mu_2(\theta)$. We demand that the solutions to these differential equations are regular functions of $\theta$. 
Solving $f_3(\theta)=0=f_4(\theta)$, we find
\begin{align}\label{eq:Dipole-Hair-l-0}
    \mu_1(\theta) &= \mu_1\,,\\
    \mu_2(\theta) &= \mu_{(2,0)} + \mu_{(2,1)}P_1(\cos(\theta))\,,\\
    \mu_{(2,0)} &= M\mu_1\,,
    \label{eq:Dipole-Hair-Formula}
\end{align}
where $\mu_1,\mu_{(2,0)}$ and $\mu_{(2,1)}$ are constants of integration which denote the $\ell=0$ mode of the monopole hair, the $\ell=0$ mode of the dipole hair and the $\ell=1$ mode of the dipole hair respectively. The function $P_1(\cdot)$ denotes the first Legendre polynomial. 
Therefore, we see that
\begin{itemize}
    \item The monopole hair $\mu_1$ is independent of $\theta$.
    \item The $\ell=0$ mode of 
the dipole hair $\mu_{(2,0)}$ is completely determined by the monopole hair and is equal to $M\mu_1$.
\end{itemize} 
Thus, the results above generalize the relation  obtained in Eq.~\eqref{eq:Dipole-hair-spherical} for spherically-symmetric spacetimes to the case of axisymmetric and circular spacetimes.
\subsubsection{Formula for the monopole hair \texorpdfstring{$\mu_1$}{Lg}}
\begin{figure}[t]
    \centering
    \includegraphics{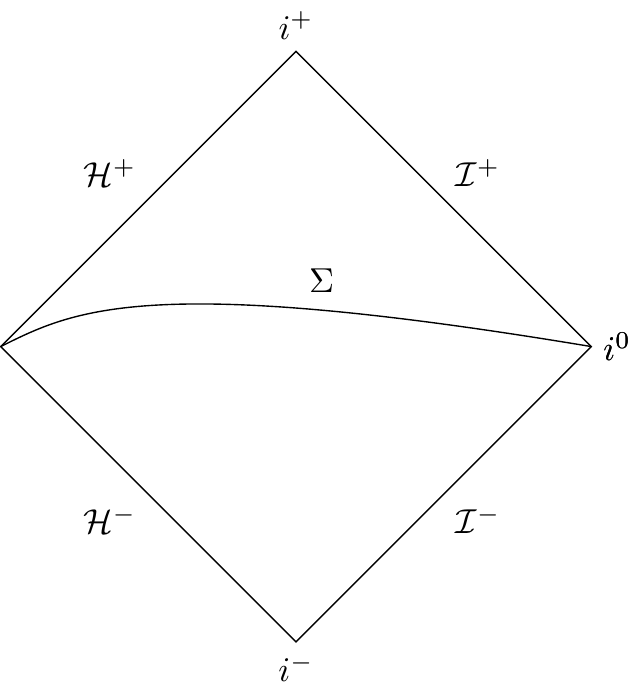}
    \caption{Penrose diagram for axisymmetric BH spacetimes. In the figure, $\mathcal{H}^{\pm}$ denote the future and the past EHs, $\mathcal{I}^{\pm}$ denote the future and the past null infinity, $i^{\pm}$ denote the future and past time like infinities, and $i^0$ denotes the point at spatial infinity. $\Sigma$ is a partial Cauchy surface.}
    \label{fig:hypersurface-lemma}
\end{figure}
We now provide a formula for the monopole hair $\mu_1$. The idea is to integrate the scalar field equation $E_{\Phi}$ and use Stokes theorem. We will also use the same technique to obtain a formula for the $\ell=1$ mode of the dipole hair. We summarise the technique in the following Lemma, which is a formal statement for applying integration by parts:
\begin{lemma}
Suppose we have an equation of the form
\begin{equation}\label{eq:Lemma-1-eq}
    \nabla_{\mu}J^{\mu} + \epsilon\, \mathcal{S} = 0\,.
\end{equation}
The EH horizon of our spacetime is a null surface generated by $X^{\mu} = t^{\mu} + \omegaH\, \phi^{\mu}$,  
where, $t^{\mu}$ and $\phi^{\mu}$ denote the Killing vectors of our spacetime, and $\omegaH$ is the angular velocity of the EH. Assume that
\begin{enumerate}
    \item $\mathcal{L}_{t}J^{\mu} = \mathcal{L}_{\phi}J^{\mu} = 0 \implies \mathcal{L}_{X} J^{\mu} = 0$,
    \item $t^{\mu}J_{\mu} = \phi^{\mu}J_{\mu} = 0 \implies X^{\mu} J_{\mu} = 0$.
\end{enumerate}
The operator $\mathcal{L}$ in the above equations denotes the Lie derivative operator.
Let $\Sigma$  be a partial Cauchy surface, as shown in Fig.~\ref{fig:hypersurface-lemma}, and let $d\Sigma_{\mu}$ be the surface element on this hypersurface. Then,
\begin{align}\label{eq:J-asymptotics}
   -\lim_{r \to \infty} \int J^{r}\sqrt{-g}\,\,d\theta d\phi 
    = {\epsilon}\int_{\Sigma} \mathcal{S} \; X^{\mu}\,d\Sigma_{\mu}\,.
\end{align}
\end{lemma}
\noindent A proof of this Lemma is provided in Appendix~\ref{Appendix:Proof-Lemma-1}. 

We now apply Lemma 1 to $E_{\Phi}$,
\begin{align}
    E_{\Phi} &= \Box{\Phi(r,\theta)}+ \epsilon \,\mathcal{F}(r,\theta) \nonumber\\
    \implies J^{\mu} &= \nabla^{\mu}\Phi \, ,\, \mathcal{S} = \mathcal{F}.
\end{align}
Since $\Phi$ respects the symmetries of the spacetime, it is easy to check that the $J^{\mu}$ defined above satisfies all the assumptions we specified in Lemma 1. To use Eq.~\eqref{eq:J-asymptotics}, we need the asymptotic expansions of
$J^r = A^{-2}\partial_r\Phi$ and $\sqrt{-g} =  r^2 A B^2 N \sin(\theta)$. Plugging in the asymptotic expansions of the metric and the scalar field, we obtain
\begin{align}
    J^r \sqrt{-g} = -\mu_1 \sin(\theta) +\mathcal{O}(r^{-1})\nonumber\,,\\
    \implies \lim_{r \to \infty} \int J^{r}\sqrt{-g}\,\,d\theta d\phi = -4\pi\mu_1\,.
\end{align}
Thus, Eq.~\eqref{eq:J-asymptotics} now gives us a formula for $\mu_1$, namely
\begin{equation}\label{eq:mu_1-Covariant}
    \mu_1 = \frac{\epsilon}{4\pi}\int_{\Sigma} \mathcal{F}(r,\theta) X^{\mu}\,d\Sigma_{\mu}\,,
\end{equation}
which is consistent with the results of~\cite{Prabhu-Stein-2018,2016-Yagi-Stein-Yunes}.

\subsubsection{Green's identity and formula for \texorpdfstring{$\mu_{(2,1)}$}{Lg}}\label{sec:Green}
Here we provide a formula for the $\ell=1$ mode of the dipole hair $\mu_{(2,1)}$. To do this, we will establish a Green's identity. Let $\phi_0$ be a stationary and axisymmetric solution of the homogeneous scalar field equation
\begin{equation}\label{eq:Hom-eq}
    \Box{\phi_0} = 0\,.
\end{equation}
We multiply $E_{\Phi}$ by $\phi_0$ and simplify it as follows
\begin{align}\label{eq:greens-identity}
    \phi_0 \Box{\Phi} +\epsilon\, \phi_0\, \mathcal{F} = 0 \nonumber\\
    \phi_0 \nabla_{\mu} \nabla^{\mu}\Phi = - \epsilon \,\phi_0\, \mathcal{F} \nonumber\\
    \nabla_{\mu}(\phi_0 \nabla^{\mu} \Phi) - \nabla^{\mu}\phi_0 \nabla_{\mu}\Phi = - \epsilon\, \phi_0\, \mathcal{F} \nonumber\\
    \nabla_{\mu}(\phi_0 \nabla^{\mu} \Phi) - \nabla_{\mu}(\Phi \nabla^{\mu} \phi_0)+ \underbrace{\Phi\,\Box{ \phi_0} }_0  = - \epsilon\, \phi_0\, \mathcal{F} \nonumber\\
    \implies \nabla_{\mu}\left[\phi_0 \nabla^{\mu} \Phi - \Phi \nabla^{\mu} \phi_0 \right] = - \epsilon\, \phi_0\, \mathcal{F} \, .
\end{align}
The equation above also satisfies all the requirements of Lemma 1 since both $\Phi$ and $\phi_0$ respect the symmetries of the spacetime. 

To apply Lemma 1 to Eq.~\eqref{eq:greens-identity}, we need to know the asymptotic properties of
\begin{equation}\label{eq:J_r-dipole}
    J^r = \left(\phi_0 \,\partial_{r} \Phi - \Phi\, \partial_{r} \phi_0\right)A^{-2}\,.
\end{equation}
We know the asymptotic expansion of $\Phi$ from Eq.~\eqref{eq:Asymptotic-Axi-Scalar}.
To determine the asymptotic properties of $\phi_0$, we start by noticing that the homogeneous scalar field equation~\eqref{eq:Hom-eq} has no asymptotically flat solutions that are regular at the EH by a no-hair theorem for massless scalar fields. We refer the reader to Sec. III of Ref.~\cite{herdeiro2018asymptotically} for a proof.
Therefore, any solution of Eq.~\eqref{eq:Hom-eq} that is regular 
at the horizon, must diverge at spatial infinity.
Since we are in an asymptotically flat spacetime, these solutions must approach the flat space solutions. To determine $\mu_{(2,1)}$, we pick $\phi_0$ which has the following boundary condition at spatial infinity
\begin{equation}\label{eq:asym-leading-phi-0}
    \lim_{r\to\infty}\frac{\phi_0(r,\theta)} {rP_1(\cos(\theta))} = 1\,.
\end{equation}

The sub-leading behaviour of $\phi_0$ can be obtained as follows. The boundary condition in the equation above means that the asymptotic expansion of $\phi_0$ has the following form
\begin{equation}
    \phi_0 = r P_1(\cos(\theta))\left[1+ \frac{\phi_1(\theta)}{r} + \mathcal{O}(r^{-2})\right]\,.
\end{equation}
We plug this equation into Eq.~\eqref{eq:Hom-eq} and use the asymptotic expansion of the metric variables in Eqs.~\eqref{eq:asym-circular-1}-\eqref{eq:asym-circular-4} to find
\begin{dmath}\label{eq:hom-asym}
    \Box{\phi_0} = \frac{f_{02}(\theta)}{r^2} +\mathcal{O}(r^{-3}),
\end{dmath}
where
\begin{align}\label{eq:f02}
    f_{02}(\theta) &= \left(1-\xi ^2\right) \xi \, \phi _1''(\xi ) \,+\, \nonumber\\
    &\left(2-4 \xi ^2\right) \phi _1'(\xi )
    -2 \xi  \phi _1(\xi ) -2 M \xi\,,
\end{align}
and $\xi = \cos(\theta)$.
Solving $f_{02}(\theta) = 0$, one obtains
\begin{equation}
    \phi_1(\theta) = -M.
\end{equation}
This gives us the following asymptotic behaviour for $\phi_0$
\begin{equation}\label{eq:asymp-hom-sol}
   \phi_0(r,\theta) = r\cos(\theta)\left[1 - \frac{M}{r} + \mathcal{O}(r^{-2})\right]\,.
\end{equation}

With this, we are now ready to understand the asymptotic expansion of $J^r$ in Eq.~\eqref{eq:J_r-dipole}. The determinant of the metric is $\sqrt{-g} = r^2 A B^2 N \sin(\theta) $. Using the asymptotic expansion of the metric variables in Eqs.~\eqref{eq:asym-circular-1}-\eqref{eq:asym-circular-4} and Eq.~\eqref{eq:asymp-hom-sol}, we see that
\begin{align}
    J^r \sqrt{-g} &= r\sin(\theta)\left[-2 \mu_1 \cos (\theta )\right] +\\ &\left[-3 \mu_{(2,1)}  \cos ^2(\theta )+2 \mu _1 M \cos (\theta )\right]\sin(\theta) \\
    &+ \mathcal{O}(r^{-1})\,.
\end{align}
Integrating over $\theta$ and $\phi$ gives
\begin{align}
    \int J^r \sqrt{-g}\,d\theta d\phi = -4\pi \mu_{(2,1)} + \Oo(r^{-1}) \nonumber,\\
    \implies -\lim_{r \to \infty} \int J^{r}\sqrt{-g}\,\,d\theta d\phi  =4\pi \mu_{(2,1)} \,.
\end{align}
Hence, applying Eq.~\eqref{eq:J-asymptotics} to the Green identity of Eq.~\eqref{eq:greens-identity}, we obtain the following formula for $\mu_{(2,1)}$,
\begin{equation}\label{eq:mu_2-1-Covariant}
    \mu_{(2,1)} = \frac{\epsilon}{4\pi} \int_{\Sigma}\mathcal{F}(r,\theta)\phi_0(r,\theta) X^{\mu}\,d\Sigma_{\mu}\,.
\end{equation}
With this, we have obtained a complete classification of the asymptotic properties of the scalar field in a large class of theories in AMC coordinates. The asymptotic expansion of the scalar field is given by,
\begin{equation}
    \Phi(r,\theta) = \frac{\mu_1}{r} + \frac{M\mu_1 + \mu_{(2,1)}P_1(\cos(\theta))}{r^2} + \Oo(r^{-3})\,,
\end{equation}
where $\mu_1$ and $\mu_{(2,1)}$ are constants independent of $\theta$. The formulas for $\mu_1$ and $\mu_{(2,1)}$ are given in Eqs.~\eqref{eq:mu_1-Covariant} and ~\eqref{eq:mu_2-1-Covariant}, respectively.

We close this section by commenting on the use of AMC coordinates.
We start by noting that we used AMC coordinate system crucially in only two places in our derivation. 
We used it first to infer the asymptotic properties of $E_{\Phi}$ to derive differential equations for $\mu_{1}(\theta)$ and $\mu_{2}(\theta)$ (see Eqs.~\eqref{eq:asymptotic-eqscalar}-\eqref{eq:f_4}). 
We then used it to derive the sub-leading behaviour of the homogeneous solution $\phi_0(r,\theta)$ in Eqs.~\eqref{eq:hom-asym}-\eqref{eq:asymp-hom-sol}. 
Therefore, the only requirement for the formula we derived to be valid is that the metric asymptotically approaches the AMC expansions given in Eqs.~\eqref{eq:asym-circular-1}-\eqref{eq:asym-circular-4}.

To repeat our calculations in any other coordinate system, the reader can essentially follow the same steps we followed, provided they know the asymptotic expansion of the metric coefficients in this new coordinate system. However, we note that gauge effects can enter into the definitions of the monopole and dipole hair in these coordinate systems.
For example, suppose that the coordinate system is not mass centered. Then, gauge effects can enter into the definition of the dipole hair.
To see this, shift $r = r_1 + a$. The asymptotic expansion of the scalar field now changes to
\begin{equation}
    \Phi(r_1,\theta) = \frac{\mu_1}{r_1} + \frac{(M\mu_1+a) + \mu_{(2,1)}P_1(\cos(\theta))}{r_1^2} + \Oo(r_1^{-3})\,.
\end{equation}
Therefore, $\mu_1$ and $\mu_{(2,1)}$ are unaffected but $\mu_{(2,0)}$ is shifted.


\subsection{Applications to Scalar Gauss-Bonnet Theory and Dynamical Chern-Simons Theory }\label{sec:App-SGB-DCS}
In this section, we apply the results obtained in the previous section to SGB theory and DCS gravity theories in the AMC Hartle-Thorne type coordinate system. We will begin by proving that the $\ell=1$ mode of the dipole hair vanishes in SGB theory. We then combine our results with that of Ref.~\cite{Prabhu-Stein-2018} to provide an exact formula for the asymptotic expansion of the scalar field to $\Oo(r^{-2})$ and relate the monopole and dipole hair to the topology of the EH.
We proceed by investigating the asymptotic expansion of scalar field in DCS theory and prove that $\mu_1^{\text{DCS}} = 0 = \mu_{(2,0)}^{\text{DCS}}$. Finally, we show how to use our formula in the decoupling limit and derive expressions for $\mu_{(2,1)}^{\text{DCS}}$ for spinning BHs.

These results imply that, during dynamical gravitational collapse in SGB theory, any angular dependence in scalar radiation has to be rapidly radiated away as we settle to a BH to \textit{all} orders in perturbation theory. For DCS theory, spherically-symmetric scalar radiation has to be rapidly radiated away in dynamical collapse to \textit{all} orders in perturbation theory. Therefore, SGB theory and DCS theory have opposite parity with respect to scalar radiation during axisymmetric dynamical collapse. 

\subsubsection{Scalar Gauss-Bonnet theory}

We now establish that the SGB theory scalar field has no $\ell=1$ dipole degree of freedom. 
The scalar field equation for SGB theory is given by
\begin{equation}\label{eq:sgb-KG}
    E_{\Phi} := \Box{\Phi(r,\theta)}+ \epsilon \,\mathcal{R}_{\text{GB}} = 0\,.
\end{equation}
By a direct calculation, one can verify that, if the metric is reflection symmetric, then the Gauss-Bonnet scalar is reflection symmetric, i.e.,
\begin{equation}
    \mathcal{R}_{\text{GB}}(r,\theta) = \mathcal{R}_{\text{GB}}(r,\pi -\theta)\,.
\end{equation}
We also see that the homogeneous solution $\phi_0(r,\theta)$ is \textit{anti-symmetric} under reflection,
\begin{equation}
    \phi_0(r,\theta) = - \phi_0(r,\pi -\theta)\,.
\end{equation}

We now use these observations in the formula for $\mu_{(2,1)}$ given in  Eq.~\eqref{eq:mu_2-1-Covariant}
\begin{equation}
    \mu_{(2,1)}^{\text{SGB}} = \frac{\epsilon_{\text{SGB}}}{4\pi} \int_{\Sigma}\mathcal{R}_{\text{GB}}(r,\theta)\phi_0(r,\theta) X^{\mu}\,d\Sigma_{\mu}\,.
\end{equation}
Let us choose $\Sigma$ to be the $t=\text{constant}$ hypersurface. This means that
\begin{align}
   X^{\mu}\,d\Sigma_{\mu} &= X^{\mu} \partial_{\mu}t\sqrt{-g} \sin(\theta)dr d\theta d\phi \nonumber\,,\\
   &= \sqrt{-g}\sin(\theta) dr d\theta d\phi
\end{align}
and
\begin{equation}
    \mu_{(2,1)}^{\text{SGB}} = \frac{\epsilon_{\text{SGB}}}{4\pi} \int_{\Sigma}\mathcal{R}_{\text{GB}}(r,\theta)\phi_0(r,\theta) \sqrt{-g}\sin(\theta) dr d\theta d\phi\,.
\end{equation}
The determinant of the metric and $\mathcal{R}_{\text{GB}}$ are even under reflection and $\phi_0(r,\theta)$ is odd under reflection. Therefore, the integrand in the equation above is odd under reflection. The integral of any function which is odd under reflections vanishes when integrated over a sphere. Thus, the integral in the equation above vanishes. This means that
\begin{equation}
    \mu_{(2,1)}^{\text{SGB}} = 0\,.
\end{equation}
Hence, the SGB scalar field has the following asymptotic behaviour to all orders in perturbation theory
\begin{equation}\label{eq:SGB-asymptotic-expansion}
    \Phi^{\text{SGB}}(r,\theta) = \frac{\mu_1^{\text{SGB}}}{r} + \frac{M\mu_1^{\text{SGB}} }{r^2} + \Oo(r^{-3})\,.
\end{equation}
This result applies to both NS and BH spacetimes.

In Ref.~\cite{Prabhu-Stein-2018} a formula was derived for the integral of the monopole hair in SGB theory at spatial infinity. To convert their coupling constant to our notation we replace $\alpha = 8\,\epsilon_{\text{SGB}}$ in their expression. The formula they derived can now be written as
\begin{equation}
    \frac{1}{4\pi}\int \mu_1^{\text{SGB}}(\theta) \sin(\theta)d\theta d\phi = 4\epsilon_{\text{SGB}}\,\kappa \,\text{Euler}(B)\,,
\end{equation}
where $\text{Euler}(B)$ denotes the Euler number of the bifurcation 2-sphere and $\kappa$ denotes the surface gravity.
In Sec.~\ref{sec:Green}, we have shown that $\mu_1(\theta)$ is independent of $\theta$ for any theory that satisfies Eq.~\eqref{eq:E-Phi}. We can thus pull $\mu_1^{\text{SGB}}(\theta) = \mu_1^{\text{SGB}} $ out of the integral to find
\begin{equation}
    \mu_1^{\text{SGB}} = 4\,\kappa\,\epsilon_{\text{SGB}}\,\text{Euler}(B)\,.
\end{equation}
Using this result in Eq.~\eqref{eq:SGB-asymptotic-expansion} gives
\begin{align}\label{eq:sgb-hair-final}
    \Phi^{\text{SGB}}(r,\theta) \!=\! \frac{ 4\,\kappa\,\epsilon_{\text{SGB}}\,\text{Euler}(B)}{r} &+ \frac{4\,M\,\kappa\,\epsilon_{\text{SGB}}\,\text{Euler}(B) }{r^2} \nonumber\\&+ \Oo(r^{-3})\,.
\end{align}
The above expression is valid to all orders in perturbation theory.
We have now obtained the monopole and dipole hair of SGB scalar field. We emphasize that without proving that $\mu_1$ is independent of $\theta$ we could not have inferred the result above. We also note that our result did not depend on perturbative arguments. For a NS spacetime, there is no bifurcation 2-sphere. Therefore,
\begin{equation}
    \Phi^{\text{SGB}}_{\text{NS}}(r,\theta) = \mathcal{O}(r^{-3})\,.
\end{equation}
We then see that both monopole and dipole hair vanish in SGB theory for a reflection symmetric NS spacetime.

To first order in perturbation theory one can substitute the GR values in Eq.~\eqref{eq:sgb-hair-final} to determine
the monopole hair and dipole hair on a BH spacetime.
The bifurcation sphere of Kerr spacetime is a 2-sphere therefore, $\text{Euler}(B) = 2$.
The surface gravity of Kerr spacetime is given by
\begin{align}
    \kappa_{\text{Kerr}} = \frac{ \sqrt{1 - \chi_{BH}^2} \left(1 - \sqrt{1 - \chi_{BH}^2} \right)}{2 M_{BH} \chi_{BH}^2}\,.
\end{align}
This expression can be found in Chapter 5.3.10 of Ref.~\cite{Poisson:2009pwt}.
Substituting these into Eq.~\eqref{eq:sgb-hair-final} we get
\begin{align}
    \mu_1^{\text{SGB},BH}&= 4 \,\kappa_{\text{Kerr}} \epsilon_{\text{SGB}} \text{Euler(B)} + \Oo({\epsilon_{\text{SGB}}^2})\,, \nonumber \\
    \label{eq:monopole-hair-sgb-kerr}
    &= \frac{4 \epsilon_{SGB}\,\sqrt{1 - \chi_{BH}^2} \left(1 - \sqrt{1 - \chi_{BH}^2} \right)}{M_{BH} \chi_{BH}^2} \nonumber\\
    &+ \Oo({\epsilon_{\text{SGB}}^2})\,. \\
    \mu_2^{\text{SGB},BH} &= M_{BH}\,\mu_1^{\text{SGB},BH} \,, \nonumber\\
    \label{eq:dipole-hair-sgb-kerr}
    &= \frac{4 \epsilon_{SGB}\sqrt{1 - \chi_{BH}^2} \left(1 - \sqrt{1 - \chi_{BH}^2} \right)}{ \chi_{BH}^2} \nonumber\\
    &+ \Oo({\epsilon_{\text{SGB}}^2})\,.
\end{align}
These results are valid for BHs of arbitrary rotation. 

\subsubsection{Dynamical Chern-Simons theory}
The scalar field equation for DCS theory in our notation is given by
\begin{equation}\label{eq:dcs-KG}
    E_{\Phi} := \Box{\Phi(r,\theta)}+ \epsilon_{\text{DCS}} \,R^{*}R = 0\,.
\end{equation}
We now apply the results we derived in the previous section to show that $\mu_{1}^{\text{DCS}} = 0$. In Ref.~\cite{Wagle:2018tyk}, it was shown that
\begin{equation}
    \frac{1}{4\pi}\int \mu_1^{\text{DCS}}(\theta) \sin(\theta)d\theta d\phi = 0\,.
\end{equation}
From Sec.~\ref{sec:Green}. we know that $\mu_1^{\text{DCS}}(\theta)$ is independent of $\theta$. Therefore,
\begin{equation}
    \mu_{1}^{\text{DCS}} = 0\,.
\end{equation}
This means that the DCS pseudoscalar has the following asymptotic expansion
\begin{equation}
    \Phi^{\text{DCS}}(r,\theta) = \frac{\mu_{(2,1)}^{\text{DCS}} P_1(\cos(\theta))}{r^2} +\Oo(r^{-3})\,
\end{equation}
where
\begin{equation}\label{eq:mu-2-1-DCS}
    \mu_{(2,1)}^{\text{DCS}} =  \frac{\epsilon_{\text{DCS}}}{4\pi} \int_{\Sigma}\left[R_{\beta\alpha\gamma\delta}{}^{*}R^{\alpha\beta\gamma\delta}\right]\phi_0(r,\theta) X^{\mu}\,d\Sigma_{\mu}\,.
\end{equation}
The formula above does not depend on perturbative arguments and, thus, it is valid to \textit{all} orders in perturbation theory. 

To illustrate how to use the formula obtained above, we derive $\mu_{(2,1)}^{\text{DCS}}$ to \textit{first} order in perturbation for arbitrarily spinning BHs.
To first order in perturbation theory, the BH background solution is just the Kerr solution. It is easy to check that Kerr metric in Boyer-Lindquist coordinates approaches the AMC Hartle-Thorne coordinate system. We will thus work in Boyer-Lindquist coordinates.
The homogeneous solution in the Kerr spacetime is given by
\begin{equation}
    \phi_0^{\text{Kerr}}(r,\theta) = (r-M)P_1(\cos(\theta))\,.
\end{equation}
The Pontryagin scalar is given by
\begin{align}
    &R{}^{*}R = \nonumber\\
    &\frac{96 M^3 r \chi  \cos (\theta ) \left(3 M^4 \chi ^4 \cos ^4(\theta )-\!10 M^2 r^2 \chi ^2 \cos ^2(\theta )+\!3 r^4\right)}{\left(M^2 \chi ^2 \cos ^2(\theta )+r^2\right)^6}
\end{align}
where $\chi = a/M$ denotes the dimensionless spin.
To use Eq.~\eqref{eq:mu-2-1-DCS} choose a $t = \text{constant}$ hypersurface so that $d\Sigma_{\mu} = \sqrt{-g_{\text{Kerr}}} \,\delta_{\mu}^{0}$. This then gives us
\begin{align}
     \mu_{(2,1)}^{\text{DCS}} =  &\frac{\epsilon_{\text{DCS}}}{4\pi} \int\left[R_{\beta\alpha\gamma\delta}{}^{*}R^{\alpha\beta\gamma\delta}\right]\phi_0^{\text{Kerr}}(r,\theta) \sqrt{-g_{\text{Kerr}}}\,dr d\theta d\phi\,\nonumber \\
     &+ \Oo(\epsilon_{\text{DCS}}^2)\,.
\end{align}
Using the expressions for the Pontryagin scalar and the homogeneous solutions in the Kerr spacetime, one can integrate the equation above to find
\begin{align}
    &\mu_{(2,1)}^{\text{DCS}}\! =\nonumber\\
    &\!\frac{2\epsilon_{\text{DCS}} \left[2 \chi ^4+\left(2 \sqrt{1-\chi ^2}-3\right) \chi ^2-2 \sqrt{1-\chi ^2}+2\right]  }{\chi ^3} \nonumber\\
    &+\Oo(\epsilon_{\text{DCS}}^2)\,,
    \label{eq:dipole-hair-dcs-kerr}
\end{align}
which is valid for BHs of arbitrary rotation. The formula derived above matches that derived in Ref.~\cite{2012-Yagi-Yunes-Tanaka-short} when one replaces $\alpha = 4\epsilon_{\text{DCS}}$ and $\beta = 2$ in their expressions.
\section{Gravitational Collapse and Scalar Radiation in SGB and DCS}\label{sec:dynamics}
Here we present slowly-rotating solutions in SGB theory and DCS theory in NS and BH spacetimes in Sec.~\ref{sec:slow_rot}, and provide a simple description of the scalar dynamics using these analytical solutions. 
Finally, in Sec.~\ref{sec:rapid_rot} we show that the analytical results presented in the previous sections match well with a dynamical numerical simulation of a rapidly rotating NS solution collapsing to a BH in the decoupling limit.

\subsection{Slowly Rotating Solutions}
\label{sec:slow_rot}
In our previous work~\cite{R:2022cwe}, we showed that the appearance of the EH and the decay of the homogeneous solution result in the growth of hair during spherically-symmetric gravitational collapse in SGB gravity.
We note that unlike the spherically-symmetric case considered in Ref.~\cite{R:2022cwe}, there are no theorems like the Kay-Wald theorem~\cite{Kay_1987} for axi-symmetric gravitational collapse.
Thus, we cannot rigorously show that the appearance of the EH and the regularity of the scalar is tied to the growth of monopole and dipole hair. 
Nevertheless, we now present arguments in favour of the same mechanism for axi-symmetric gravitational collapse in SGB theory and DCS theory by analyzing stationary NS and BH solutions in the slow-rotation approximation.
We provide numerical evidence confirming these results in Sec.~\ref{sec:rapid_rot} for a rapidly rotating NS collapsing to a BH.

\subsubsection{Scalar Gauss-Bonnet theory}
On a slowly rotating NS spacetime, the Gauss-Bonnet scalar field is given by
\begin{equation}
    \Phi^{\text{SGB}}_{\text{NS}} = \epsilon_{\text{SGB}}\Phi_0(r) + \mathcal{O}(\chi^2)\,, 
\end{equation}
where
\begin{align}
    \Phi_0(r) &= \frac{2}{M_{\text{NS}}\,r} + \frac{2}{r^2} + \frac{8M_{\text{NS}}}{3r^3} + \frac{1}{M_{\text{NS}}^2}{\log \left(1 - \frac{2M_{\text{NS}}}{r} \right)}\,.
\end{align}
The above expression falls of as $r^{-4}$ asymptotically.
As we noted in Ref.~\cite{R:2022cwe}, the appearance of the EH and the radiation of the homogeneous solution results in the growth of hair during dynamical gravitational collapse from a NS spacetime to a BH spacetime.
The scalar monopole and dipole hair around the newly-formed, rotating BH is determined by Eq.~\eqref{eq:monopole-hair-sgb-kerr} and \eqref{eq:dipole-hair-sgb-kerr}. 
Assuming reflection symmetry, we showed in Eq.~\eqref{eq:SGB-asymptotic-expansion} that the dipole hair of the SGB scalar field is given by
\begin{equation}
    \mu_2 = M \mu_1\,.
\end{equation}
Therefore, the structure of dipole radiation must be very similar to the structure of monopole scalar radiation. Moreover, this implies that the appearance of the EH and the decay of the homogeneous part of the solution must result in the growth of both dipole hair and monopole hair at late times.
\begin{figure*}
    \centering
    \includegraphics[width=\textwidth]{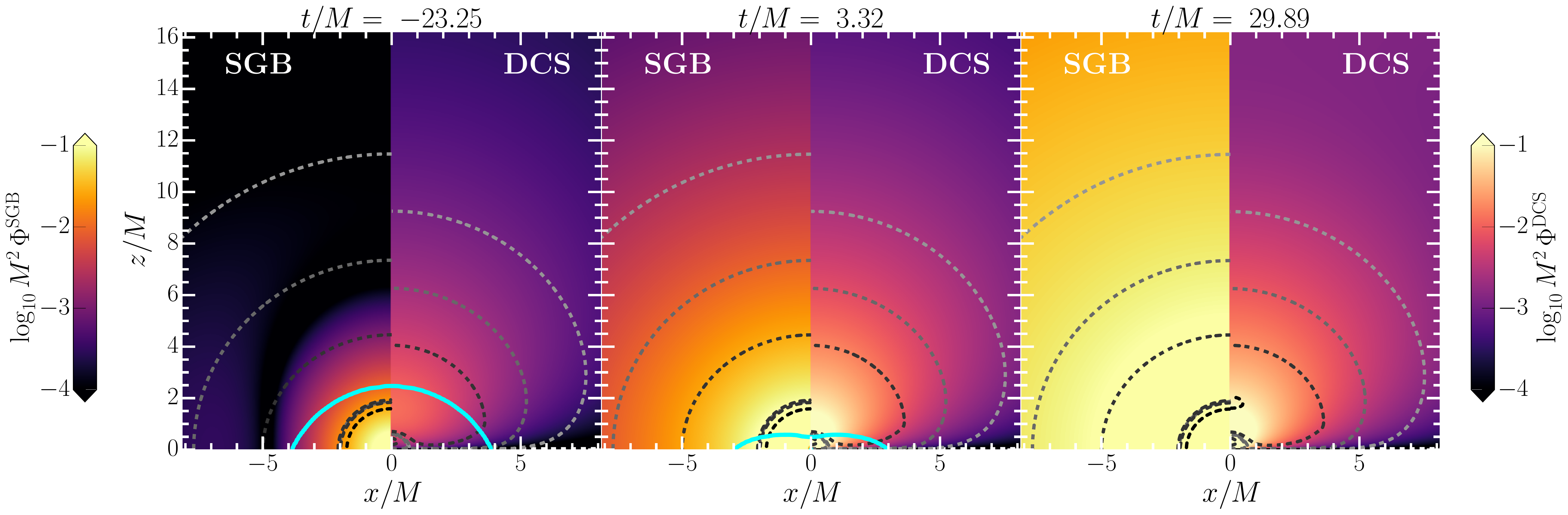}
    \caption{Dynamics of the scalar field $\Phi$ in SGB (left) and DCS (right) theories of gravity during the collapse of a rapidly rotating star. The cyan line denotes the surface  of the star. The contours denote 
    of the Gauss-Bonnet invariant $\mathcal{R}_{\rm GB}$ and of Pontryagin density $R^{*}R$, respectively.
    Specifically, the isocurvature contours are $[10^{-3},10^{-4},10^{-5},10^{-6},10^{-9}] M^{-4}$ for each theory, respectively.
    All units are stated relative to the initial mass of the star, and all times are stated relative to the time of AH formation at $t=0$.}
    \label{fig:NR_collapse}
\end{figure*}
\subsubsection{Dynamical Chern-Simons theory}
On a slowly-rotating NS spacetime, the profile of the DCS scalar field is given by~\cite{Ali_Ha_moud_2011}
\begin{align}
        &\Phi^{\text{DCS}}_{\text{NS}}(r,\theta) = \nonumber\\
        &\epsilon_{\text{DCS}}  P_1(\cos(\theta))\left\{\frac{J_{NS} \left(18 M_{\text{NS}}^2+10 M_{\text{NS}} r+5 r^2\right)}{2 M_{\text{NS}}^2 r^4} \right. \nonumber\\
        &- \left. \frac{5 C_1 J_{NS} \left((M_{\text{NS}}-r) \log \left(1-\frac{2 M_{\text{NS}}}{r}\right)-2 M_{\text{NS}}\right)}{4 M_{\text{NS}}^5} \right\} + \nonumber \\
        &\mathcal{O}(\chi^3) \,. \label{eq:DCSNS}
\end{align}
Above, the constant $C_1$ multiplying the homogeneous solution is obtained by matching the exterior to the interior solution at the surface of the star~\cite{Ali_Ha_moud_2011}.
On a BH background, the homogeneous part of the solution is not present and the profile is given by
\begin{align}
&\Phi^{\text{DCS}}_{\text{BH}}(r,\theta) = \nonumber\\
&\epsilon_{\text{DCS}} P_1(\cos(\theta))\left\{\frac{J_{\text{BH}}  \left(18 M_{\text{BH}}^2+10 M_{\text{BH}} r+5 r^2\right)}{2 M_{\text{BH}}^2 r^4}\right\} \nonumber\\
&+ \mathcal{O}(\chi^3).
\label{eq:DCSBH}
\end{align}
Thus, if the solution has to be regular during dynamical collapse, the homogeneous solution has to radiated away. We note that monopole hair is absent for both NS and BH solutions in DCS.
This means that the $\ell=0$ part of the dipole radiation must also be absent during dynamical collapse apart from transients.
Hence, scalar radiation for the DCS scalar field is strongest along the z-axis and must fall-off as $r^{-2}$.

\subsection{Dynamical Collapse of Rapidly Rotating Solutions}
\label{sec:rapid_rot}

In this section, we present numerical relativity simulations of the dynamical collapse of rapidly-rotating neutron stars. 
These simulations not only go beyond the limitations of the 
slow-rotation approximation presented in Sec. \ref{sec:slow_rot}, but they also provide numerical confirmation of the results presented in this work.

To this end, we extend our previous simulations of spherical collapse in perturbative SGB gravity \cite{R:2022cwe} to rapidly rotating stars.
More specifically, we numerically solve a dynamical GR background on top of which we evolve the decoupled Klein-Gordon equations for the DCS theory \eqref{eq:dcs-KG} and SGB theory \eqref{eq:sgb-KG} cases. 
This is done within the 3+1 split of the four dimensional spacetime \cite{Arnowitt:1962hi}, identifying a set of spacelike hypersurfaces $\left(\Sigma_t,\gamma_{ij}\right)$ with a time coordinate $t$ and induced spatial three-metric $\gamma_{ij}$.
Within this decomposition, we evolve the conservation of matter and energy-momentum equations for a perfect fluid \cite{Font:1998hf}.

The initial data for a rapidly rotating neutron star spacetime is constructed using the \texttt{RNS} code \cite{Stergioulas:1994ea}, implementing the method outlined in \cite{Komatsu:1989zz}.
We then use this code to construct the rotating neutron star model similar to D4 of \cite{Baiotti:2004wn}. We will briefly describe it's properties in the following.
With a rotation frequency of $f \simeq 1300\, \rm Hz$ and a polar to equatorial axis ratio of $0.65$, this is one of the fastest rotating configurations we can construct.
Since we are not interested in the nuclear physics aspects of the star, we choose a simple $\Gamma=2$ ideal-fluid, a relation where the specific internal energy  $\varepsilon$, the pressure $p$, and rest-mass density $\rho$ of the fluid are related by $p = \rho \varepsilon \left(\Gamma -1\right)$. The initial values for the internal energy density are constructed using a polytrope $\varepsilon \left(t=0\right) = K \rho^{\left(\Gamma-1\right)}/\left(\Gamma -1\right)$ with $K=100$. This results in a neutron star mass of $1.86\, M_\odot$ and a dimensionless spin $\chi = J/M^2 = 0.54$, where $J$ is the angular momentum of the initial star.

Initially, we set the scalar field to zero, but evolve the star for a short time for the hair to grow a steady-state solution. We remark that, although the star is dynamically unstable under any form of perturbation, these instabilities grow slowly enough that hair can still form before the star begins to collapse in earnest. We then accelerate collapse by adding a small inwards pointed velocity perturbation to the star.

We perform the numerical evolution using the \texttt{Einstein Toolkit} infrastructure \cite{yosef_zlochower_2022_6588641,Loffler:2011ay}. 
In detail, we solve the Einstein equations in the Z4c formulation with moving puncture gauges \cite{Hilditch:2012fp,Bernuzzi:2009ex}, together with the equations of general-relativistic ideal magnetohydrodynamics \cite{Duez:2005sf}. Numerically, these are implemented in the  \texttt{FIL} code \cite{Most:2019kfe}, which is derived from the publicly available \texttt{IllinoisGRMHD} code \cite{Etienne:2015cea}. \texttt{FIL} employs a formally fourth-order accurate numerical method for both the matter \cite{DelZanna:2007pk} and spacetime sectors \cite{Zlochower:2005bj}, which has been demonstrated to be third-order accurate for matter spacetimes \cite{Most:2019kfe}.  Following \cite{Okounkova:2017yby,Witek:2018dmd}, we have extended \texttt{FIL} to evolve the decoupled scalar field equations for SGB theory~\eqref{eq:sgb-KG} and DCS theory~\eqref{eq:dcs-KG} spacetimes. The code has been tested against the publicly available \texttt{CANUDA} code \cite{witek_helvi_2021_5520862}. More details will be presented in a forthcoming work.

The numerical grid is constructed using a set of nested rectangular boxes \cite{Schnetter:2003rb} at doubling resolution. Our finest resolution is $\Delta x = 0.086\, M$ with a total number of five nested boxes. 
Starting from the onset of collapse we compute the location of the apparent and event horizons using the \texttt{AHFinderDirect} \cite{Thornburg:2003sf} code and our own implementation of the algorithm presented in \cite{Diener:2003jc}, respectively.
While the problem is intrinsically two-dimensional, we compute the problem in all three dimensions, with reflection symmetry applied across the equatorial plane of the initial star.

\begin{figure}[b!]
    \centering
    \includegraphics[width=0.5\textwidth]{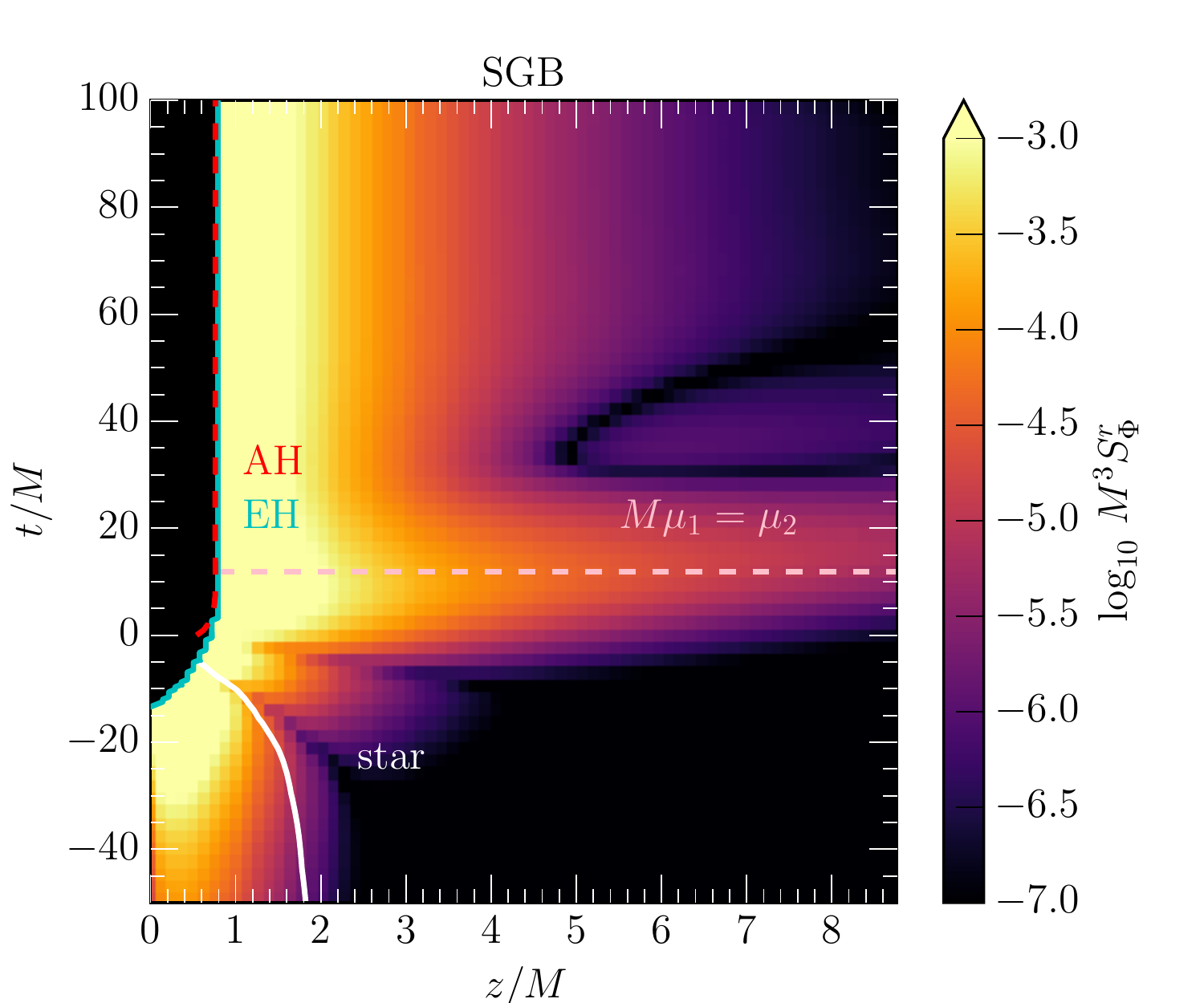}\\
    \includegraphics[width=0.5\textwidth]{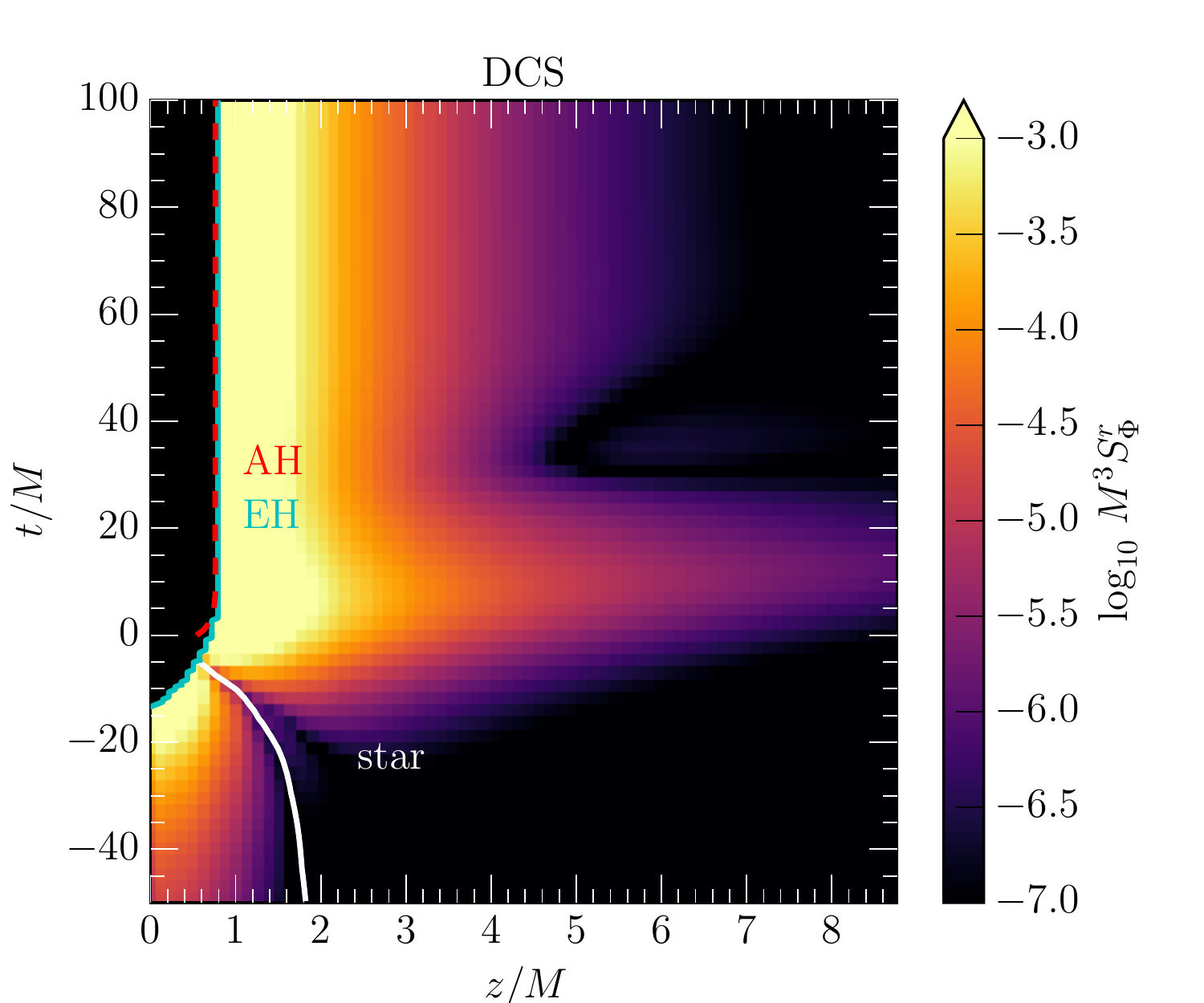}
    \caption{Evolution of the scalar field radial energy flux, $S^r_\Phi$, for the SGB theory (top) and DCS theory (bottom) spacetimes.
    Shown is the time evolution along the rotation axis of the star, which coincides with the z-axis of the domain.
    We further highlight the surface of the star (white), the EH (cyan) and the AH (red). The interior of the black hole has been masked out. For the case of SGB gravity, we also mark the time at which the monopolar ($\mu_1$) and dipolar ($\mu_2$) charges match the analytic prescription (pink). All times $t$ are stated relative to the time of AH formation.}
    \label{fig:NR_flux}
\end{figure}

\begin{figure*}
    \centering
    \includegraphics[width=0.7 \textwidth]{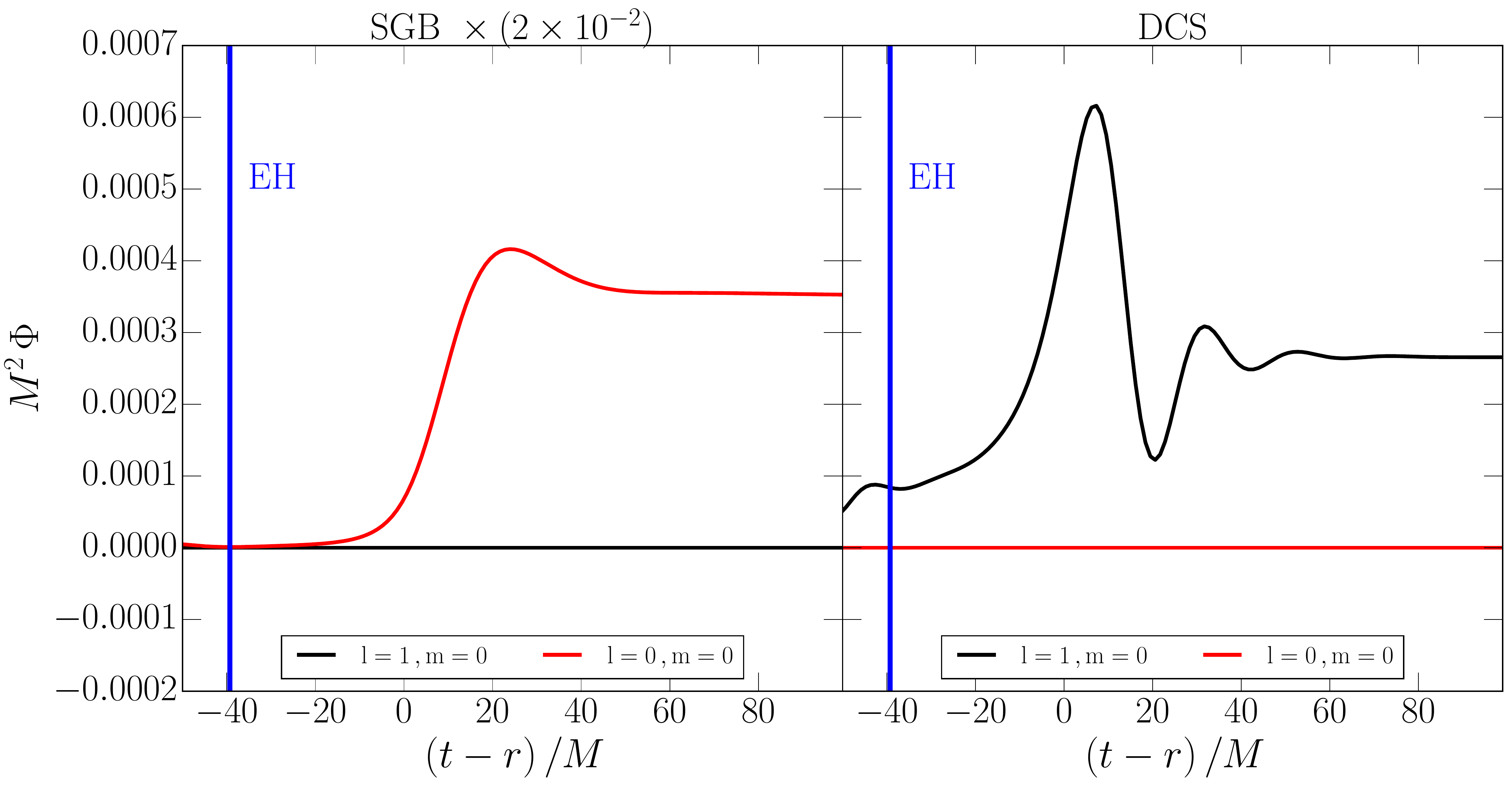}
    \caption{Evolution of the $\ell=0$ (red) and $\ell=1$ (black) modes of the scalar hair, 
    shifted in time by the extraction radius $r=27\, $M.
    For SGB gravity, only the $\ell=0$ mode grows. For DCS, after an initial transient during collapse, a stable $\ell=1$ mode has developed as predicted by Eq.\ \eqref{eq:DCSBH}.
    For reference we also show the mode that vanishes due the parity of the respective field.
    The formation time of the black hole is indicated by the time an EH is first found.}
    \label{fig:NR_l1}
\end{figure*}

\begin{figure*}
    \centering
    \includegraphics[width=0.7\textwidth]{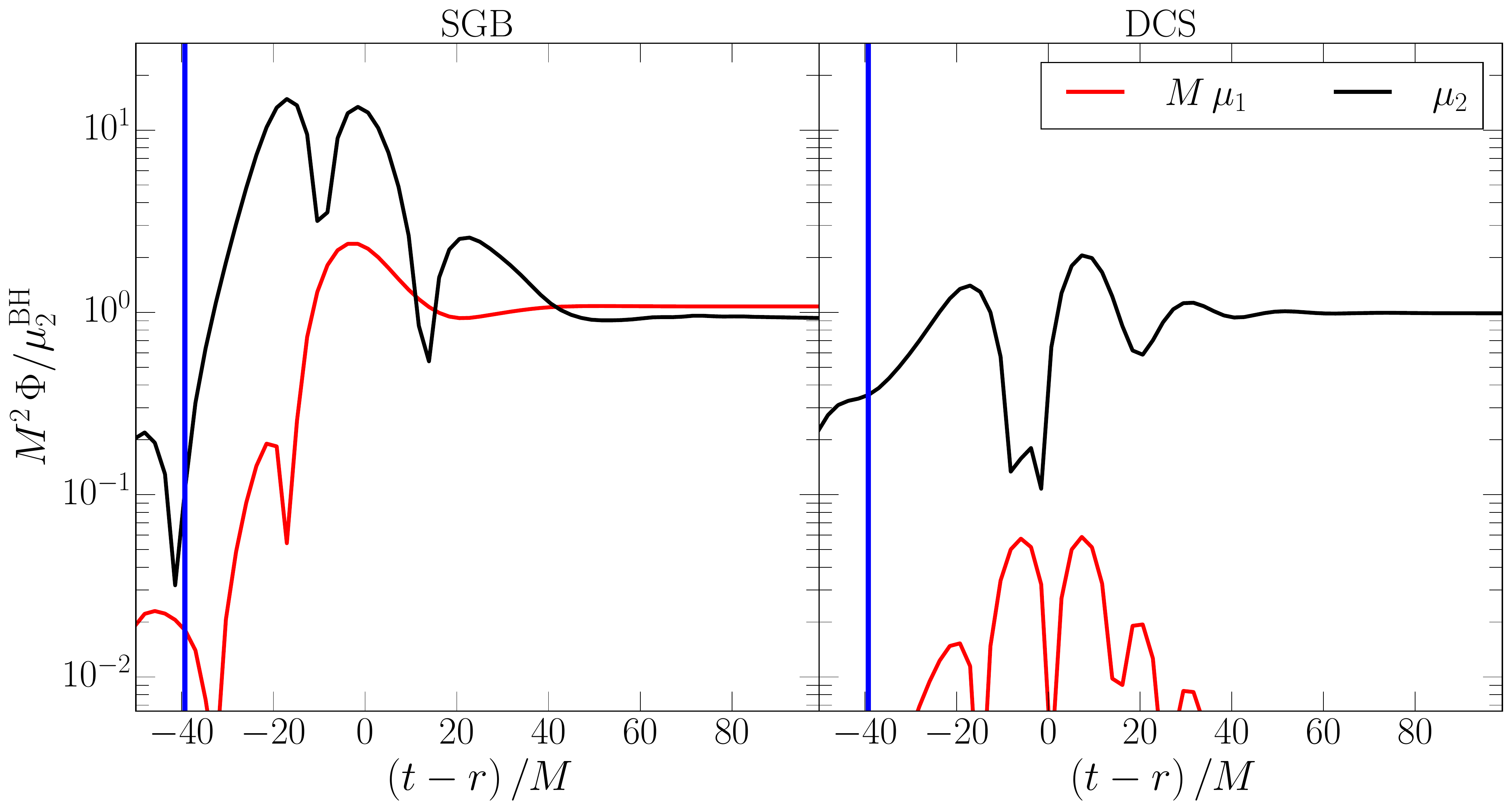}
    \caption{Evolution of monopolar ($\mu_1$) and dipolar ($\mu_2$) hair, normalized by the final values of $\mu_2^{\rm BH}$ for the resulting black holes in each theory, see Eq. \eqref{eq:dipole-hair-sgb-kerr} and \eqref{eq:dipole-hair-dcs-kerr}.
     Shown is the evolution of the monopolar and dipolar hair charges in the equatorial plane (SGB) and along the pole (DCS), normalized by the Komar mass, $M$, of the initial star in order to match Eq.\ \eqref{eq:Dipole-Hair-Formula}. For SGB initially no dipolar hair is present it grows to the analytically determined value after the collapse. For DCS the monopolar hair grows during the formation but is radiated away as soon as the final black hole rings down.
     The initial formation time of the black hole is indicated by the time an EH is first found (blue lines). We state all times relative to $r=27\, M$.}
    \label{fig:NR_sGB}
\end{figure*}
We begin by describing the overall dynamics of the collapse. 
The dynamics of the matter fields in this case has been discussed extensively (e.g., \cite{Shibata:2003iw,Baiotti:2004wn}), and will not be repeated here. In short, matter on the rotational axis falls in first (due to lack of rotational support), whereas matter on the equator remains outside of the BH longest. Massive disks are not formed in this process \cite{Margalit:2015qza,Camelio:2018gfc} so that the mass of the BH and of the initial star will approximately coincide, with the difference being given by gravitational wave emission \cite{Stark:1985da,Baiotti:2007np}.

Since the dynamics of matter has been discussed extensively elsewhere, let us instead focus on the description of the scalar field dynamics. Figure \ref{fig:NR_collapse} shows the scalar fields $\Phi_{\rm SGB/DCS}$ in SGB and DCS gravity, respectively. Starting from the left, we see that the scalar field attains values of $\Phi M^2 < 10^{-4} \ (10^{-2})$ before collapse in the SGB (DCS) gravity case.
During collapse, the scalar field in the SGB theory case begins to grow monopole and dipole hair, leading to a strong growth of the field by a factor of 100, compared to the initial field strength. We will discuss this in more detail in the following paragraphs. For the DCS case, the growth is less pronounced, which is consistent with the absence of monopole hair in this theory (see Eq.~\eqref{eq:DCSBH}).

In Paper I \cite{R:2022cwe}, we performed an in-depth analysis of when the hair begins to develop in the case of a spherically-symmetric (non-rotating) star. Here, we extend this analysis to the case of rapid rotation. Since this aspect of the discussion is nearly identical for both SGB gravity and DCS theory spacetimes, we only focus on the latter. We refer the reader to Paper I \cite{R:2022cwe} for further details on the setup of this analysis. Figure\ \ref{fig:NR_flux} shows the radial scalar field energy flux, $S^r_\Phi$, along the rotation axis of the star. This axis is most convenient, as the pseudo-scalar DCS field vanishes on the equator. We then track the collapse of the star (indicated by a white line) into the BH (black area). Due to the symmetry properties along the rotation axis, the EH finding problem \cite{Diener:2003jc} reduces to a one-dimensional problem, which we solve in post-processing. We also show the apparent horizon (AH), computed in full three dimensions \cite{Thornburg:2003sf}, for reference, with both agreeing at late times, as expected. 
We can now see that, different from the non-rotating case, a scalar field flux is present already when the star begins to contract. The EH is only formed when the star has already collapsed substantially, i.e., at $t\approx 140 M$. However, the strongest energy flux of the scalar field occurs only after horizon formation.

Finally, let us perform a quantitative analysis to assess the validity of the analytic results obtained in Sec. \ref{sec:slow_rot} for slowly rotating NS. In short, these concern the growth of an $\ell=1$ mode in the DCS pseudo-scalar field (see \eqref{eq:DCSBH}), as well as the presence of dipole hair in SGB, which is related to monopole hair \eqref{eq:Dipole-Hair-Formula} only by the mass of the system.
We can address both of these questions by studying properties of the scalar field at \textit{intermediate} 
distances $(r=10-40\, M)$ from the BH. We begin by considering the DCS case.
Before collapse, the field $\Phi^{\rm DCS}$ sourced by the rotating neutron star will have an $\ell=1$ mode only (see Eq.~\eqref{eq:DCSBH}). After collapse the BH retains this mode exclusively, with only the magnitude of the $\ell=1$ component changing to Eq.~\eqref{eq:DCSBH}. We find that this behavior holds for rapidly rotating NS, as well. Figure\ \ref{fig:NR_l1} shows the $\ell=1$ mode of the DCS pseudo-scalar field, $\Phi^{\rm DCS}$. As expected, the amplitude of the pseudo-scalar field grows during collapse, with the $\ell=0$ mode being absent also during the transient phase around $t-r\sim 90\, M$.
We therefore conclude that the hair before, during, and after collapse will only have an $\ell=1$ mode in the small coupling limit of DCS gravity for slowly and rapidly spinning NS.

In the case of SGB gravity, our main finding concerns the growth of dipole hair and the simple relation between monopole and dipole charges (see Eq.~\eqref{eq:Dipole-Hair-Formula}). 
Although only established in the static case, we confirm these findings numerically for the dynamical case. To this end we compute the scalar field profile over a range of radii $r=[5.5; 27] M$, and perform a quadratic fit in order to compute the $\mu_1$ and $\mu_2$ charges during the entire time of the collapse. The resulting evolution is shown in 
Fig.~\ref{fig:NR_sGB}. We find that dipole hair grows during collapse, and, after an initial transient around $t=100-150\, M$, it attains its expected value at late times.

\section{Conclusions and Future Directions}\label{sec:Conclusions}
In this paper, we have investigated the dynamics of scalar monopole and dipole radiation in a wide range of theories in axi-symmetric spacetimes.
In particular, we have shown in Eq.~\eqref{eq:Dipole-Hair-Formula} that a simple relation exists between the $\ell=0$ mode of the scalar dipole hair and the scalar monopole hair for a wide class of theories in axi-symmetric and reflection symmetric spacetimes.
We then used this result to study two specific modified theories, i.e., SGB gravity and DCS theory in spherical and axi-symmetric spacetimes.

In Paper-I we studied the dynamics of scalar monopole radiation in SGB theory in spherically-symmetric spacetimes. Our dipole hair formula allowed us to directly translate the result from Paper-I to dipole hair in SGB theory. 
In that paper, we found that monopole
hair in SGB theory grows during gravitational collapse, as a NS collapses to form a BH. 
Our result [Eq.~\eqref{eq:Dipole-Hair-Formula}] shows that the dipole hair should also grow in spherically symmetric collapse.
We then extended these results to axisymmetric gravitational collapse in both SGB and DCS theories. 
By working in the slow-rotation approximation, we showed that the DCS scalar field solution has divergent modes that have to radiated away if the scalar field is to remain regular during gravitational collapse on and exterior to the EH.
This result shows that mechanism responsible for growth of hair during gravitational collapse in SGB  and DCS are exactly the same, albeit the parity of scalar radiation is opposite in these two theories.

Finally, we confirmed our analytical predictions using numerical simulations of gravitational collapse of a rapidly rotating NS star in the decoupling limit.
Our results show that the appearance of the EH results in strong scalar radiation, which results in the growth of scalar monopole and dipole hair in SGB and the growth dipole hair in DCS.
Therefore, the results of this paper and Ref.~\cite{R:2022cwe} provide a complete picture of the far-field dynamics of scalar radiation in a wide class of theories, including SGB theory and DCS gravity.

Our results present some natural directions for future investigations.
One possible direction is to see how our results change when we go beyond the decoupling analysis.
So far, for the theories considered in this paper, a locally well-posed initial value formulation only exists for SGB theory
when deviations from GR are ``small'' (see~\cite{Kovacs:2020pns}).
The analytical results of Sec.~\ref{sec:Monopole-Dipole}
did not make any assumptions about the decoupling limit.
The numerical results, more particularly, the emission of strong scalar radiation and its correlation with the appearance of the EH, did use the decoupling assumption.
When the coupling constant is small, we expect our analysis to carry over if one includes the back reaction of the scalar field onto the metric.
It would be interesting to confirm this prediction in the future.

Binary BH collisions have been simulated in SGB theory~\cite{Witek:2018dmd,Okounkova:2019zep,Ripley-East-2021,CCZ4-2022} and in DCS gravity~\cite{Okounkova:2017yby}.
Binary NS collisions have also been simulated in SGB theory recently~\cite{East-2022}, using the modified harmonic formulation~\cite{Kovacs:2020pns}.
One direction for future work involves using the analytical results established in this paper to understand the dynamics of scalar radiation better in SGB theory and DCS theory.
Another natural direction would be to establish general results for gravitational radiation and leading-order metric corrections to GR.
Understanding the dynamical behaviour of the leading-order metric corrections will help build a theoretical understanding complementing the post-Newtonian approximation~\cite{2012-Yagi-Stein-Yunes-Tanaka,Shiralilou:2020gah,Shiralilou:2021mfl} and numerical studies.

Finally, one could also consider more general theories than the ones we considered [Eq.~\eqref{eq:Action}] and see which of these results carry over.
The interesting aspects of theories such as SGB gravity is that the monopole hair vanishes for neutron star spacetimes, but the monopole hair is not zero for a BH spacetime~\cite{Prabhu-Stein-2018}.
It would be interesting to see if the same results hold in other theories; some examples of which could be theories with a more general coupling function $f(\phi)$ to the curvature scalar $\F$~\cite{Prabhu-Stein-2018} or actions that contain higher powers of the Riemann curvature, which naturally arise in a gradient expansion around GR.
If this happens, then, one could study the growth of scalar hair in these theories and see if the growth is related to the emission of scalar radiation from the appearance of the EH.
It would also be interesting to see if formula such as the one we derived in Eq.~\eqref{eq:Dipole-Hair-Formula} hold in these theories.
\begin{acknowledgements}
AH and NY acknowledge support from the Simons Foundation through Award number 896696.
ERM acknowledges support as John A. Wheeler Fellow at the Princeton Center for Theoretical Science, as well as postdoctoral fellowships at the Princeton Gravity Initiative, and the Institute for Advanced Study. JN is partially supported by the U.S. Department of Energy, Office of Science, Office for Nuclear Physics under Award No. DE-SC0021301. 
HW acknowledges support provided by NSF grants No. OAC-2004879 and No. PHY-2110416, and Royal Society (UK) Research Grant RGF\textbackslash R1\textbackslash 180073.

This work used the Extreme Science and Engineering Discovery Environment (XSEDE),
through the allocation TG-PHY210114, 
which was supported by NSF grants No.~ACI-1548562 and No.~PHY-210074.
This research used resources provided by the Delta research computing project, which is supported by the NSF Award No. OCI~2005572 and the State of Illinois. Delta is a joint effort of the University of Illinois at Urbana-Champaign and its National Center for Supercomputing Applications.
The authors acknowledge the Texas Advanced Computing Center (TACC) at The University of Texas at Austin for providing HPC resources that have contributed to the research results reported within this paper, under LRAC grants AT21006.
Part of the simulations presented in this article were performed on computational resources managed and supported by Princeton Research Computing, a consortium of groups including the Princeton Institute for Computational Science and Engineering (PICSciE) and the Office of Information Technology's High Performance Computing Center and Visualization Laboratory at Princeton University.
\end{acknowledgements}
\input{Appendix}
\bibliographystyle{apsrev4-1}
\bibliography{ref}
\end{document}

%% file: Appendix.tex
\appendix
\section{Constructing Asymptotically Mass Centered Coordinates}
In this appendix we prove that one can always construct an AMC coordinate system suitable for our purposes in this paper.
\subsection{Static And Spherically Symmetric Spacetimes}\label{appendix:AMC-spherical}
The proof for spherically symmetric and static spacetimes follows by performing simple translations. In an ingoing null coordinate system $x^{\mu} = (v,r,\theta,\phi)$, the metric is given by,
\begin{equation}
    ds^2 = -D(r)dv^2 + 2dvdr + K(r) d\Omega^2\,.
\end{equation}
The asymptotic expansions of $D(r)$ and $K(r)$ are given by
\begin{align}
    D(r) = 1- \frac{2M}{r} + \Oo(r^{-2})\,,\\
    K(r) = r^2\left( 1 - \frac{K_{1}}{r} + \Oo(r^{-2})\right)\,,
\end{align}
where, as before, $M$ denotes the Komar mass of the spacetime. If $K_1$ is equal to zero then we are in AMC coordinates. If $K_1$ is not equal to zero then shift, $r = R + K_1$. It is easy to see that in the coordinate system $x^{\mu} = (v,R,\theta,\phi)$ the asymptotic expansion of the metric functions is given by
\begin{align}
    D(R) = 1- \frac{2M}{R} + \Oo(R^{-2})\\
    K(R) = R^2\left( 1 + \Oo(R^{-2})\right)\,.
\end{align}
Thus, the coordinate system $x^{\mu} = (v,R,\theta,\phi)$ is AMC.

\subsection{Stationary, Axisymmetric, Circular And Reflection Symmetric Spacetimes}\label{appendix:AMC-circular}
In case of stationary, axisymmetric, and circular spacetimes one can always introduce Hartle-Thorne type coordinates~\cite{1967-Hartle-Sharp}. The line element is given by
\begin{align}\label{eq:Line-HT}
    ds^2 &= -N^2(r,\theta)\, dt^2 + A^2(r,\theta)\, dr^2 \, + \nonumber\\
     &r^2B^2(r,\theta) \left\{ d\theta^2 + \sin(\theta)^2 \left[d\phi - \omega(r,\theta) dt\right]^2\right\}\,.
\end{align}
To prove that we can express these coordinates in AMC form we need to show that the asymptotic expansions of the metric functions, $N(r,\theta),\, A(r,\theta),\, B(r,\theta)$, and $\omega(r,\theta)$ are given by Eqs.~\eqref{eq:asym-circular-1}-\eqref{eq:asym-circular-4}. To obtain the asymptotic expansions of the metric functions we will look at the gravitational equations of motion,
\begin{equation}
    E_{\mu\nu} =  G_{\mu\nu} + 16\pi \,\epsilon\, \mathcal{C}_{{\mu\nu}} - 8\pi \left(T_{\mu\nu}^{\Phi} - T_{\mu\nu}^{\text{matter}} \right) = 0\,
\end{equation}
where the components of the tensor $E_{\mu\nu}$ are defined with respect to the tetrad
\begin{equation}
    \left\{\partial_t,\partial_r,\frac{\partial_{\theta}}{r}, \frac{\partial_{\phi}}{r\sin(\theta)}\right\}\,.
\end{equation}
We will assume that the matter stress energy tensor has compact support.
Then, in the asymptotic region $T_{\mu\nu}^{\text{matter}} = 0$. 
We further assume that the curvature scalar $\mathcal{F}$ falls off faster than $\Oo(r^{-5})$. 
The tensor $C_{\mu\nu}$~\eqref{eq:C-Tensor} is constructed from the curvature scalar $\mathcal{F}$ and the scalar field $\Phi$. Therefore, it must fall off at least as fast as $\mathcal{F}$, asymptotically. With this observation, the components of the gravitational field equations in the tetrad set up above have the form
\begin{equation}
    E_{\mu\nu} = G_{\mu\nu}  - 8\pi T_{\mu\nu}^{\Phi} + \Oo(r^{-5})\,.
\end{equation}
We also note that from the definition of $T_{\mu\nu}^{\Phi}$~\eqref{eq:T-phi} we see that $T_{\mu\nu}^{\Phi}$ falls off as $\Oo(r^{-4})$.
Thus, the field equations reduce to
\begin{equation}
    E_{\mu\nu} = G_{\mu\nu} +  \Oo(r^{-4})\,.
\end{equation}
We now substitute the following expansions into the field equations
\begin{align}
    N(r,\theta) = 1 + \sum_{j=1}^{\infty}\frac{N_{j}(\theta)}{r^j}\,,\\
    A(r,\theta) = 1 + \sum_{j=1}^{\infty}\frac{A_{j}(\theta)}{r^j}\,,\\
    \omega(r,\theta) = \sum_{j=1}^{\infty}\frac{\omega_{j}(\theta)}{r^j}\,,\\
    \Phi(r,\theta) = \sum_{j=1}^{\infty}\frac{\mu_{j}(\theta)}{r^j}\,,\\
    B(r,\theta) = \sum_{j=1}^{\infty}\frac{B_{j}(\theta)}{r^j}\,.
\end{align}
To prove that the coordinates are AMC we have to show that
\begin{align}
    \omega_1(\theta) &= 0 \,,\\
    \omega_2(\theta) &= 0\,,\\
    N_1(\theta) &= -M\,, \\
    A_1(\theta) &= M\,,\\
    \omega_3(\theta) &= 2J\,,\\
    B_1(\theta) &= 0\,.
\end{align}
We start by showing that $\omega_1(\theta) = 0$. To $\Oo(r^{-2})$ one finds
\begin{align}
    &E_{33} = -\frac{3 \sin ^2(\theta ) \left(\omega _1'(\theta ){}^2+\omega _1(\theta ){}^2\right)}{4 r^2} + \Oo(r^{-3})\,\nonumber,\\
    \implies &\omega _1(\theta ) = 0\,.
\end{align}
Next, we show that $\omega_2(\theta) = 0$. Using, $\omega_1(\theta) = 0$ we can simplify $E_{03}$ as
\begin{align}
    E_{03} &= \frac{3 \cos (\theta ) \omega _2'(\theta )+\sin (\theta ) \omega _2''(\theta )-2 \omega _2(\theta ) \sin (\theta )}{2 r^3} \nonumber\\
    &+ \Oo(r^{-4})\,.
\end{align}
The only solution to the above differential equation which is regular in $\theta$ is $\omega_2(\theta) = 0$.

We now show that $\omega_3(\theta)$ is a constant. We first notice that
\begin{align}
    T_{03}^{\Phi}& = \frac{\omega _1(\theta ) \sin (\theta ) \left(\mu _1'(\theta ){}^2+\mu _1(\theta ){}^2\right)}{4 r^4} + \Oo(r^{-5}) \nonumber\\
    &\sim\Oo(r^{-5})\,,
\end{align}
since $\omega_1(\theta) = 0$. So,
\begin{align}
    E_{03} &= G_{03} + \Oo(r^{-5})\nonumber \,, \\
    &= \frac{3 \cos (\theta ) \omega _3'(\theta )+\sin (\theta ) \omega _3''(\theta )}{2 r^4} + \Oo(r^{-5})\,.
\end{align}
The solution to the above equation which is regular in $\theta$ is a constant solution. Let us denote this constant by $\omega_3(\theta) = 2J$\,. We now derive a constraint between $A_{1},\,B_{1}$, and $N_1$. To $\Oo(r^{-3})$
\begin{align}
    E_{12} = \frac{A_1'(\theta )+B_1'(\theta )+2 N_1'(\theta )}{r^3} + \Oo(r^{-4})\nonumber\\
    \label{eq:B1-relation}
    \implies B_1(\theta) = c_0 - 2N_1(\theta) - A_1(\theta)\,.
\end{align}
Using the above constraint relation we find that,
\begin{align}
    E_{00} = -\frac{2 \left(\cot (\theta ) N_1'(\theta )+N_1''(\theta )\right)}{r^3} + \Oo(r^{-4})\,.
\end{align}
Regularity in $\theta$ implies that $N_1(\theta) = -M$. We will identify $M$ with the Komar mass of the spacetime later. Using this relation in $E_{22}$ we obtain
\begin{equation}
    E_{22} = \frac{\cot (\theta ) A_1'(\theta )+A_1(\theta )-M}{r^3} + \Oo(r^{-4})\,.
\end{equation}
The solution of the above differential equation is given by
\begin{equation}
    A_1(\theta) = M + c_1\cos(\theta)\,.
\end{equation}
We set the constant $c_1$ to zero because of the assumption of reflection symmetry. Thus,
\begin{equation}\label{eq:A1}
    A_1(\theta) = M\,.
\end{equation}
Combining the equation above with Eq.~\eqref{eq:B1-relation} we see that
\begin{equation}
    B_1(\theta) = c_0 + 2M - M = c_0 + M\,.
\end{equation}
We summarize the results we have so far below
\begin{align}
     \omega_1(\theta) &= 0\,, \\
    \omega_2(\theta) &= 0\,,\\
    N_1(\theta) &= -M \,,\\
    A_1(\theta) &= M\,,\\
    \omega_3(\theta) &= 2J\,,\\
    B_1(\theta) &= c_0 + M\,.
\end{align}
The last step now, is to get rid of the constant $c_0$.
To do this, let us shift $r = r' - c_2$, where the aim is to use the constant $c_2$ to set $c_0$ to zero. 
The line element~\eqref{eq:Line-HT} in the shifted coordinate is given by
\begin{align}
    ds^2 &= -N^2(r'-c_2,\theta)\, dt^2 + A^2(r'-c_2,\theta)\, dr'{}^2 \, + \nonumber\\
     &r'{}^2\left(1-\frac{c_2}{r'}\right)^2B^2(r'-c_2,\theta) \left\{ d\theta^2\,+ \nonumber\right.\\
     &\left.  \sin(\theta)^2 \left[d\phi - \omega(r,\theta) dt\right]^2\right\}\,.
\end{align}
Therefore, the metric function $B(r,\theta)$ transforms as
\begin{align}
    B'(r',\theta) & = \left(1-\frac{c_2}{r'}\right)^2 B(r'-c_2,\theta)\,\nonumber\\
    &= \left(1 - \frac{2c_2}{r'}\right) \left(1 + \frac{c_0 + M}{r'}\right) + \Oo(r'{}^{(-2)})\,.
\end{align}
We can set $c_0$ to zero by choosing, ${2c_2 = c_0 + M}$. This means that in the new shifted coordinate system
\begin{equation}
    B'(r',\theta) = 1 + \Oo(r'{}^{-2})\,.
\end{equation}
The leading behaviour of the other metric functions is not affected. We now drop the superscript on $r'$. Hence, we have shown that we can install a coordinate system where the metric functions have the following asymptotic behaviour
\begin{align}
     \omega_1(\theta) &= 0\,, \\
    \omega_2(\theta) &= 0\,,\\
    N_1(\theta) &= -M \,,\\
    A_1(\theta) &= M\,,\\
    \omega_3(\theta) &= 2J\,,\\
    B_1(\theta) &= 0\,.
\end{align}
One can now compute the Komar mass and angular momentum and verify that they are indeed equal to $M$ and $J$, respectively. Therefore, we have established that one can always install AMC coordinates for spacetimes respecting the assumptions (1)-(4)  made in Sec.~\ref{sec:dipole-axisymmetric}.
\section{Proof of Lemma 1}\label{Appendix:Proof-Lemma-1}
In this appendix we provide a proof of Lemma 1. The statement and proof of Lemma 1 is as follows.
\begin{namedtheorem}[Lemma 1]
Suppose we have an equation of the form
\begin{equation}
    \nabla_{\mu}J^{\mu} + \epsilon\, \mathcal{S} = 0\,.
\end{equation}
The EH horizon of our spacetime is a null surface generated by $X^{\mu} = t^{\mu} + \omegaH\, \phi^{\mu}$ where $\omegaH$ is the angular velocity of the EH. Assume that
\begin{enumerate}
    \item $\mathcal{L}_{t}J = \mathcal{L}_{\phi}J = 0 \implies \mathcal{L}_{X} J = 0$,
    \item $t^{\mu}J_{\mu} = \phi^{\mu}J_{\mu} = 0 \implies X^{\mu} J_{\mu} = 0$.
\end{enumerate}
The operator $\mathcal{L}$ in the  equations above denotes the Lie derivative operator.
Let $\Sigma$  be a partial Cauchy surface as shown in Fig.~\ref{fig:hypersurface-lemma} and let $d\Sigma_{\mu}$ be the surface element on this hypersurface. Then
\begin{align}
   -\lim_{r \to \infty} \int J^{r}\sqrt{-g}\,\,d\theta d\phi 
    = {\epsilon}\int_{\Sigma} \mathcal{S} X^{\mu}\,d\Sigma_{\mu}\,.
\end{align}
\begin{proof}
We start by defining an anti-symmetric tensor
\begin{equation}\label{eq:Q-def}
    Q^{\mu\nu} := J^{\mu}X^{\nu}-X^{\mu}J^{\nu}.
\end{equation}
Taking the divergence of $Q^{\mu\nu}$ we obtain,
\begin{equation}
    \nabla_{\nu} Q^{\mu\nu} \!= \!J^{\mu} \nabla_{\nu}X^{\nu}\! + X^{\nu}\nabla_{\nu}J^{\mu}\! - J^{\nu}\nabla_{\nu}X^{\mu} - X^{\mu}\nabla_{\nu}J^{\nu}\!.
\end{equation}
The first term in the equation above is zero because $X^{\nu}$ is a Killing vector. The second and the third term can be combined to give
\begin{equation}
    X^{\nu}\nabla_{\nu}J^{\mu} - J^{\nu}\nabla_{\nu}X^{\mu} = \mathcal{L}_{X}J^{\nu} = 0\,.
\end{equation}
Therefore, the last term is the only non-zero term, which can be simplified using Eq.~\eqref{eq:Lemma-1-eq}
\begin{equation}\label{eq:Gauss-Law-Q}
   \nabla_{\nu} Q^{\mu\nu} = - X^{\mu}\nabla_{\nu}J^{\nu}= \epsilon\, X^{\mu} \mathcal{S} \,.
\end{equation}
We now integrate the equation above on a partial Cauchy hypersurface $\Sigma$ as shown in Fig.~\ref{fig:hypersurface-lemma}. Let $d\Sigma_{\mu}$ represent the volume element on the hypersurface $\Sigma$
\begin{equation}\label{eq:Lemma-1}
    \int_{\Sigma} \nabla_{\nu}Q^{\mu\nu}d\Sigma_{\mu} = {\epsilon}\int_{\Sigma} \mathcal{S} X^{\mu}\,d\Sigma_{\mu}.
\end{equation}
The left-hand side of the above equation can be simplified by using Stokes theorem
\begin{dmath}
    \int_{\Sigma} \nabla_{\nu}Q^{\mu\nu}d\Sigma_{\mu} = \frac{1}{2}\int_{\partial\Sigma} Q^{\mu\nu}dS_{\mu\nu}\,.
\end{dmath}
The boundary $\partial \Sigma$ consists of a cross section of the EH and spatial infinity as shown in Fig.~\ref{fig:hypersurface-lemma},
\begin{equation}\label{eq:Lemma-1-1}
     \int_{\Sigma} \nabla_{\nu}Q^{\mu\nu}d\Sigma_{\mu} = \frac{1}{2}\int_{\infty} Q^{\mu\nu}dS_{\mu\nu}^{\infty} + \frac{1}{2}\int_{\mathcal{H}} Q^{\mu\nu}dS_{\mu\nu}^{\mathcal{H}}. 
\end{equation}
The surface element of the EH is given by
\begin{equation}
    dS_{\mu\nu}^{\mathcal{H}} = 2X_{\left[\mu\right.}l_{\left.\nu\right]}\sqrt{\sigma_{\mathcal{H}}}\,d\theta d\phi\,,
\end{equation}
where $l^{\mu}$ is the second null normal to the EH and $\sigma_{\mathcal{H}}$ is the determinant of the induced metric on the EH. From the definition of $Q^{\mu\nu}$~[Eq.~\eqref{eq:Q-def}] we see that
\begin{align}
    \left.Q^{\mu\nu}X_{\left[\mu\right.}l_{\left.\nu\right]}\right|_{\mathcal{H}} &= \left.Q^{\mu\nu}X_{\mu}l_{\nu}\right|_{\mathcal{H}} \nonumber\\
    &= \left.\left(X_{\mu}J^{\mu} X^{\nu} - X_{\mu}X^{\mu}J^{\nu}\right)\right|_{\mathcal{H}} = 0\,.
\end{align}
The first term is zero by assumption and the second term is zero because $X^{\mu}$ is null on the horizon. Thus, the integral over the EH is zero
\begin{align}
   \frac{1}{2}\int_{\mathcal{H}} Q^{\mu\nu}dS_{\mu\nu}^{\mathcal{H}} &= \int_{\mathcal{H}}\left.Q^{\mu\nu}X_{\left[\mu\right.}l_{\left.\nu\right]}\right|_{\mathcal{H}}\sqrt{\sigma_{\mathcal{H}}}\,d\theta d\phi\nonumber\\
   &= \left. \int_{\mathcal{H}}Q^{\mu\nu}X_{\mu}l_{\nu}\right|_{\mathcal{H}} \sqrt{\sigma_{\mathcal{H}}}\,d\theta d\phi= 0\,.  
\end{align}
Therefore, Eq.~\eqref{eq:Lemma-1-1} simplifies to
\begin{equation}
    \int_{\Sigma} \nabla_{\nu}Q^{\mu\nu}d\Sigma_{\mu} = \frac{1}{2}\int_{\infty} Q^{\mu\nu}dS_{\mu\nu}^{\infty} \,.
\end{equation}
The surface element at spatial infinity is given by
\begin{equation}
    dS_{\mu\nu}^{\infty} = 2\partial_{\left[\mu\right.}t\,\partial_{\left.\nu\right]}r\sqrt{-g}\,d\theta d\phi\,.
\end{equation}
We note that at spatial infinity, $X^{\mu}\partial_{\mu}t = 1$ and $X^{\mu}\partial_{\mu}r $ evaluates to zero everywhere. Hence
\begin{align}
     &\int_{\Sigma} \nabla_{\nu}Q^{\mu\nu}d\Sigma_{\mu} = \frac{1}{2}\int_{\infty} Q^{\mu\nu}dS_{\mu\nu}^{\infty} \nonumber\\
     &=\int_{\infty}Q^{\mu\nu}\partial_{\left[\mu\right.}t\,\partial_{\left.\nu\right]}r\sqrt{-g}\,d\theta d\phi\nonumber\\
     &= \int_{\infty}Q^{\mu\nu}\partial_{\mu}t\partial_{\nu}r\sqrt{-g}\,d\theta d\phi\nonumber\\
     &=\int_{\infty}\left(\underbrace{J^{\mu}\partial_{\mu}t}_{0}\underbrace{X^{\nu}\partial_{\nu}r}_{0} - J^{\nu}\partial_{\nu}r\underbrace{X^{\mu}\partial_{\mu}t}_{1}\right)\sqrt{-g}\,d\theta d\phi \nonumber\\
     &= -\int_{\infty} J^r\sqrt{-g}\,d\theta d\phi := -\lim_{r \to \infty} \int J^{r}\sqrt{-g}\,\,d\theta d\phi \,.
\end{align}
Comparing the equation above with Eq.~\eqref{eq:Lemma-1} we obtain the result we intended.
\end{proof}
\end{namedtheorem}